\documentclass[12pt,a4paper]{article}
\usepackage{amsmath,amssymb}
\usepackage{amsfonts}
\usepackage{graphicx}
\usepackage{theorem}
\usepackage{color}
\usepackage[font={scriptsize,sf,sl}]{caption,subfig}
\newtheorem{theorem}{Theorem}
\newtheorem{lema}{Lemma}
\newtheorem{corollary}{Corollary}
\newtheorem{proposition}{Proposition}
\newtheorem{definition}[theorem]{Definition}

\def \ba {\begin{array}}
\def \ea {\end{array}}

\def \bea {\begin{eqnarray}}
\def \eea {\end{eqnarray}}

\def \be {\begin{equation}}
\def \ee {\end{equation}}

\def\nn{\nonumber}

\begin{document}

\begin{center}
{\LARGE {\bf Wavelet transform on the torus: a group theoretical approach}}
\end{center}
\bigskip
\bigskip

\centerline{{\sc Manuel Calixto}$^{1}$, {\sc Julio Guerrero}$^{2}$ and {\sc Daniela Ro\c sca}$^{3}$ }

\bigskip

\begin{center}

{\it $^1$ Department of Applied Mathematics, University of Granada, Faculty of Sciences, Campus de Fuentenueva, 18071 Granada, Spain }
\\
{\it $^2$ Department of Applied Mathematics, University of Murcia,  Faculty of Informatics, Campus de Espinardo,
30100 Murcia, Spain}
\\
{\it $^3$ Department of Mathematics, Technical University of Cluj-Napoca, str. Memorandumului 28, RO-400114, Cluj-Napoca, Romania
}

\end{center}

\bigskip
\begin{center}
{\bf Abstract}
\end{center}
\small
\begin{list}{}{\setlength{\leftmargin}{3pc}\setlength{\rightmargin}{3pc}}
\item We construct a Continuous Wavelet Transform (CWT) on the torus $\mathbb T^2$ following a group-theoretical approach 
based on the conformal group $SO(2,2)$. The Euclidean limit reproduces  wavelets on the plane $\mathbb R^2$ with two dilations, which 
can be defined through the natural tensor product representation of usual wavelets on $\mathbb R$. Restricting ourselves to a single 
dilation imposes severe conditions for the mother wavelet that can be overcome by adding  extra modular group $SL(2,\mathbb Z)$ 
transformations, thus leading to the concept of \emph{modular wavelets}. We define modular-admissible functions and prove frame conditions.

\end{list}
\normalsize 



\section{Introduction}

The original idea of Jean Baptiste Joseph Fourier on the possibility of decomposing a given function
into a sum of sinusoids, basic ``waves'' or ``harmonics'',  has  exerted an enormous influence upon science and engineering. Since its beginnings, Harmonic Analysis 
has been developed with the goal of explaining a wide range of physical phenomena in diverse
fields as: Optics, x-ray Crystallography, 
Computerized Tomography,  Nuclear Magnetic Resonance, Radioastronomy and Modern Cosmology, 
 and, at a more mathematical (fundamental) level,
Number Theory, Diophantine Equations, Riemann zeta function, Ergodic Theory, Probability Theory, 
Automorphic Functions,  etc.  Last, but not least, Harmonic Analysis is deeply rooted in the foundations of Quantum Mechanics.

Large sections of some of these subjects may be looked upon as nearly identical with certain branches of
the theory of group representations. Actually, it was Hermann Weyl and Fritz Peter in 1927 who pointed
out and emphasized the (still insufficiently appreciated) fact that
classical Fourier analysis can be illuminatingly regarded as a chapter in
the representation theory of compact commutative Lie groups.

Nowadays, perhaps one of the most successful and popular applications of Harmonic Analysis is the \textit{Theory of Wavelets}, which has become an
important branch of numerical and applied mathematics, sharing with Approximation Theory the search of expansions in terms of 
functions belonging to more accessible functional spaces  due to their structural characteristics
and their computational simplicity (viz, polynomial, splines, rational functions, etc). However, we must say that the wavelet 
idea was already rooted in Quantum Mechanics under the more 
general notion of \textit{coherent state}.  The term ``coherent'' itself originates
in the current language of quantum optics (for instance, coherent
radiation). It was introduced in the 1960s by Glauber and it was Aslaksen and Klauder who first studied the
one-dimensional affine group, for the purely quantum mechanical purpose of
generalizing the standard uncertainty relations ``position-momentum'' (or
time-frequency), for the Heisenberg group, to ``dilation-translation'' . It
was yet another mathematical physicist, Alex Grossmann, 
who discovered the crucial link between the representations of the affine
group and the intriguing technique in signal analysis developed by Jean Morlet. 

Since the pioneer work of Grossmann, Morlet and Paul
\cite{GMP}, several extensions of the standard Continuous Wavelet Transform (CWT) on $\mathbb R$
to general manifolds $\mathbb X$ have been constructed (see e.g.
\cite{Gazeau,Fuhr} for general reviews and
\cite{CWTmanifolds,Fuhr2} for recent papers on WT and Gabor
systems on homogeneous manifolds). Particular interesting examples
are the construction of CWT on: spheres $\mathbb{S}^{N-1}$, by
means of an appropriate unitary representation of the Lorentz
group  in $N+1$ dimensions $SO(N,1)$
\cite{Holschneider-sphere,waveS2,nsphere,Sphere-Implementation,AVdiscreteS2}, on the upper sheet
$\mathbb H^2_+$  of the two-sheeted hyperboloid $\mathbb H^2$
\cite{cwthyperbol}, or its stereographical projection onto the
open unit disk $D_1=SO(1,2)/SO(2)$, and the construction of conformal wavelets in the (compactified) 
complex Minkowski space \cite{MacMahon}.  
The basic ingredient in all these constructions is
a group of transformations $G$ which contains dilations and motions on
$\mathbb X$, together with a transitive action of $G$ on $\mathbb X$.

In this article we first extend the group theoretical construction of wavelets on the circle $\mathbb S^1$ based on the group $SL(2,\mathbb R)$, given in \cite{wavecircle}, to 
wavelets on the  two-torus $\mathbb T^2=\mathbb S^1\times \mathbb S^1$ based on the group $SO(2,2)$, and introduce additional modular transformations 
in $SL(2,\mathbb Z)$, which lead to the concept of {\it modular wavelets}. 

We must stress that the topological 
torus $\mathbb T^2=(\mathbb R/2\pi\mathbb Z)^2$ can be obtained from the plane $\mathbb R^2$ 
by imposing periodic boundary conditions and these are often 
used in physical and mathematical models to simulate a large system by modeling a small part that is far from its edge. 
For instance, in the Quantum Hall Effect \cite{Prange}, the topology of the problem is that of a torus \cite{CMP}, and modular transformations
are of crucial importance for the classification of fractional quantum numbers \cite{modular}.
Moreover, the Discrete Fourier Transform, either in one or more dimensions, implicitly assumes that the signal or image is 
periodic, and this is a valid approximation as long as edge effects are negligible. Besides,  wavelets on $\mathbb{R}^2$ (or higher dimensions) encounters applications in microlocal 
analysis \cite{waveletmicro}, and thus wavelets on the torus would be helpful in toroidal microlocal analysis \cite{toroidalmicro}.




The organization of the paper is as follows.  In Section \ref{secs} we briefly remind the group theoretical construction of the CWT on $\mathbb S^2$ based on the Lorentz group 
$SO(3,1)$, which serves as an introduction 
and to set notation. In Section \ref{sect2} we construct the CWT on the topological torus $\mathbb T^2$ based on the group $SO(2,2)$, 
introducing admissibility conditions and proving the existence of admissible functions and continuous wavelets frames. This construction  
naturally relies on two dilations. Usual wavelet constructions rely on a single dilation but, in our construction, the frame property is lost 
when restricting to a single (let us say, diagonal) dilation. The way out is to introduce additional ingredients in the wavelet parameter space, like 
{\it modular transformations}, which lead to the concept of {\it modular wavelets}. This construction is made in Section \ref{secmodular}.

\section{CWT on the sphere $\mathbb S^2$ based on $SO(3,1)$: a reminder}\label{secs}

Let us denote by $L^2(\mathbb S^2,d\Omega)$ the Hilbert space of square integrable functions
on the two-sphere $\mathbb S^2$, with the usual measure $d\Omega=\sin\theta d\theta d\varphi$ 
(we shall omit $d\Omega$ and just write $L^2(\mathbb S^2)$). An orthonormal basis of $L^2(\mathbb S^2)$ is given in terms of
spherical harmonics:
\begin{equation}
 Y_l^m(\theta,\varphi)=N_{lm}P_l^m(\cos\theta)e^{im\varphi},\;\; l=0,1, \dots,\;\; m=-l,\dots,l\label{sph}
\end{equation}
fulfilling
\begin{equation}
 \langle Y_l^m| Y_{l'}^{m'}\rangle=\int_{\theta=0}^\pi\int_{\varphi=-\pi}^\pi  \overline{Y_l^m(\theta,\varphi)} Y_{l'}^{m'}(\theta,\varphi)d\Omega=\delta_{ll'}\delta_{mm'},
\end{equation}
with a convenient choice of normalization factors $N_{lm}$, where $P_l^m$ are the associated Legendre polynomials.

The problem of defining a satisfactory dilation on the sphere was solved by Antoine and Vandergheynst
in \cite{waveS2}, where they used a group-theoretical approach based on the Lorentz group $G=SO(3,1)$. Dilation is embedded into
$G$ via the Iwasawa decomposition $G=KAN$ with $K$ compact, $A$ Abelian and $N$ nilpotent subgroups. The parameter space $X$ of their
CWT is the quotient $G/N$. The expression for the dilation, with parameter $a>0$, of the colatitude angle $\theta$ is

\begin{equation}
 \theta_a= 2\arctan(a\tan(\theta/2)),\label{dilS2angle}
\end{equation}
and it has a direct geometrical interpretation as a dilation around the North Pole of the sphere, lifted from the tangent plane by
inverse stereographic projection. For any function $f\in L^2(\mathbb S^2)$, a unitary representation of this
dilation is given by

\begin{equation}
[D_a^{\mathbb S^2}f](\theta,\varphi)=\lambda(a,\theta)^{1/2}f(\theta_{1/a},\varphi),\label{dilS2}
\end{equation}
where
\begin{equation}
 \lambda(a,\theta)= \frac{d\cos\theta_{1/a}}{d\cos\theta}  =\frac{4a^2}{((a^2-1)\cos\theta+a^2+1)^2}   \label{multisphere}
\end{equation}
is a multiplier (Radon-Nikodym derivative).
We can write points of $X$
as pairs $(\beta,a)$ with $\beta\in SO(3)$ (rotations) and $a\in (0,\infty)$ (dilations).
Given a function
$f\in L^2(\mathbb S^2)$,  the representation
\begin{equation}
 f_{\beta,a}(\theta,\varphi):= [U_\beta^{\mathbb S^2}\circ
 D_a^{\mathbb S^2}f](\theta,\varphi)\label{dilrotS2}
\end{equation}
is unitary,
where $[U_\beta^{\mathbb S^2} f](\theta,\varphi)=f(\beta^{-1}(\theta,\varphi))$ is the quasi-regular representation of
$SO(3)$.

\begin{definition}
A non-zero function $f\in L^2(\mathbb S^2)$ is called admissible iff the condition
\begin{equation}
0<\int_X d\nu(\beta,a)|\langle f_{\beta,a}|\psi\rangle|^2< \infty
\end{equation}
is satisfied for any $\psi \in L^2(\mathbb S^2)$, where $d\nu(\beta,a)=\frac{da}{a^3}d\mu(\beta)$ is the
measure on $X$ and $d\mu(\beta)$ is
the Haar measure on $SO(3)$.
\end{definition}
This also means that the representation \eqref{dilrotS2} is square integrable. A weaker (necessary but not sufficient)
admissibility condition is (see \cite{waveS2})
\begin{equation}
 \int_{\mathbb S^2} \frac{f(\theta,\varphi)}{1+\cos\theta} d\Omega=0.
\end{equation}
Given an admissible function $f\in L^2(\mathbb S^2)$, the family $\{f_{\beta,a},\beta\in SO(3), a>0\}$ is called a frame iff there exist
two real positive constants $A\leq B$ such that
\begin{equation}
 A\|\psi\|^2\leq \int_Xd\nu(\beta,a)|\langle f_{\beta,a}|\psi\rangle|^2 \leq B\|\psi\|^2, \ \forall
\psi \in L^2(\mathbb S^2).\label{frameS2}
\end{equation}

It is known that any admissible function $\phi\in L^2(\mathbb R^2)$ provides an admissible function on the sphere by
inverse stereographic projection
\begin{equation}
 [\Pi^{-1}_{\mathbb S^2} \phi](\theta,\varphi)=\frac{2 \phi(2\tan(\theta/2),\varphi)}{1+\cos\theta}.
\end{equation}

\section{CWT on the torus $\mathbb T^2$ based on the group $SO(2,2)$}\label{sect2}

Let us consider now the Hilbert space $L^2(\mathbb T^2,d\omega)$ of square integrable functions on the torus $\mathbb T^2$, with measure 
$d\omega=d\theta_1d\theta_2$, where $\theta_1, \theta_2$
are angles parametrizing the corresponding ``meridional'' and ``equatorial'' circles, respectively. This measure
is invariant under translations $\theta_{1,2}\to\theta_{1,2}+\vartheta_{1,2}$ on the torus, and arises naturally from the Haar measure
on the group $SO(2,2)$. 
We denote by $\langle \cdot|\cdot \rangle$ the inner product with respect to this measure, i.e.
$$
\langle f|g \rangle:=\int_{-\pi}^\pi\int_{-\pi}^\pi\overline{f(\theta_1,\theta_2)}{g(\theta_1,\theta_2)}d\omega,
$$
for all $f,g\in L^2(\mathbb T^2)$ (we shall omit $d\omega$ in $L^2(\mathbb T^2,d\omega)$ from now on). 
An orthonormal basis of $L^2(\mathbb T^2)$ is given in terms of ``plane wave'' functions
\begin{equation}
 \phi_{n_1n_2}(\theta_1,\theta_2)=\frac{1}{2\pi}e^{in_1\theta_1}e^{in_2\theta_2},\;\;
n_1,n_2\in\mathbb Z; \;\; \langle \phi_{n_1,n_2}|\phi_{n'_1,n'_2} \rangle=\delta_{n_1,n'_1}\delta_{n_2,n'_2}.\label{ONBtorus}
\end{equation}
The coefficients $\widehat{f}^{n_1,n_2}:=\langle \phi_{n_1,n_2}|f\rangle$ are the usual Fourier coefficients
of $f\in L^2(\mathbb T^2)$.

\subsection{The group-theoretical construction}
Again, the problem of defining a satisfactory dilation on the torus can be addressed in a group theoretical setting by
resorting to the group $SO(2,2)$, which is locally isomorphic to the direct product $SO(2,1)\times SO(2,1)$. In fact
$$SO(2,2)=(SO(2,1)\times SO(2,1))/\mathbb{Z}_2.$$
While in the case of the Lorentz
group $SO(3,1)$, the Iwasawa decomposition $KAN$ leads to a one-dimensional dilation group, in the case of $SO(2,2)$, the
Iwasawa decomposition gives a two-dimensional dilation group\footnote{The dimension of $A$ in the $G=KAN$ decomposition equals the so called (real) rank of the group $G$, which for $SO(m,n)$ is min$(m,n)$,
see \cite{Barut}, pag. 127.}. More precisely, since $SO(2,1)$ is locally isomorphic to $SL(2,\mathbb{R})$, and any $2\times2$ matrix of determinant one can be
decomposed as
\begin{equation}
\left(\begin{array}{cc} \cos(\vartheta/2) & \sin(\vartheta/2) \\ -\sin(\vartheta/2) & \cos(\vartheta/2) \end{array}\right)
\left(\begin{array}{cc} \sqrt{a} & 0 \\ 0 & 1/\sqrt{a} \end{array}\right)
\left(\begin{array}{cc} 1 & b \\ 0 & 1 \end{array}\right)\,,
\end{equation}
the $KAN$ decomposition of $SL(2,\mathbb R)$ is given by $K_1=\mathbb{T}^1=\mathbb S^1$, $A_1=(0,\infty)$ and $N_1=\mathbb{R}$. Since $SO(2,2)$ is locally the direct product of
two copies of
$SO(2,1)$, the parameter space of the CWT is now  $X=KAN/N=\mathbb{T}^2\times(0,\infty)^2$ whose points are labeled by
$(\vartheta_1,\vartheta_2,a_1,a_2)$, with
$\vartheta_i\in (-\pi,\pi)$, $a_i\in  (0,\infty)$ for $i=1,2$.

From the group law, one can see that the action of the dilation group $A$ on the torus $K$ is given by the expression

\begin{equation}
 \theta_a= 2\arctan(a\tan(\theta/2)),\;\; \theta=\theta_k,\; a=a_k,\; k=1,2.\label{stereodiltoruseq}
\end{equation}

\begin{figure}
\begin{center}
\includegraphics[height=4cm]{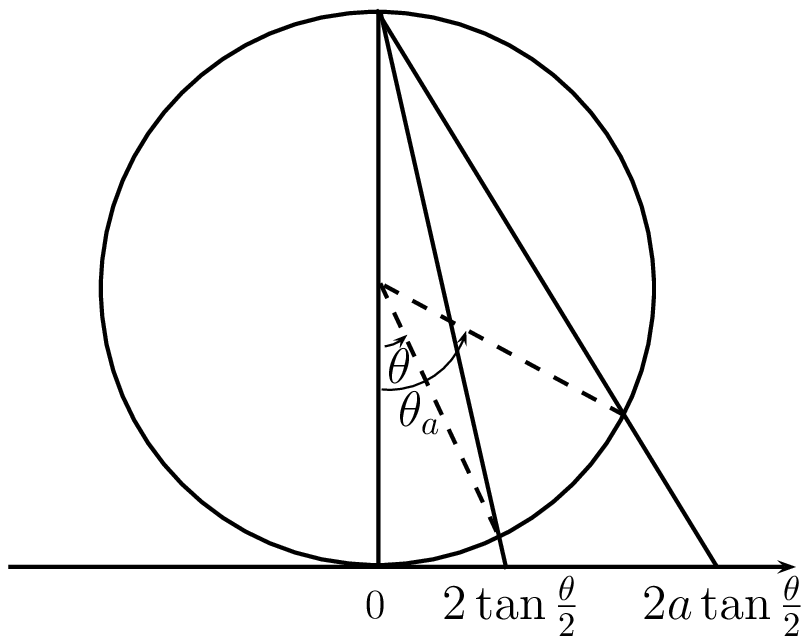}\hspace{5mm}\includegraphics[height=4.5cm,width=5cm]{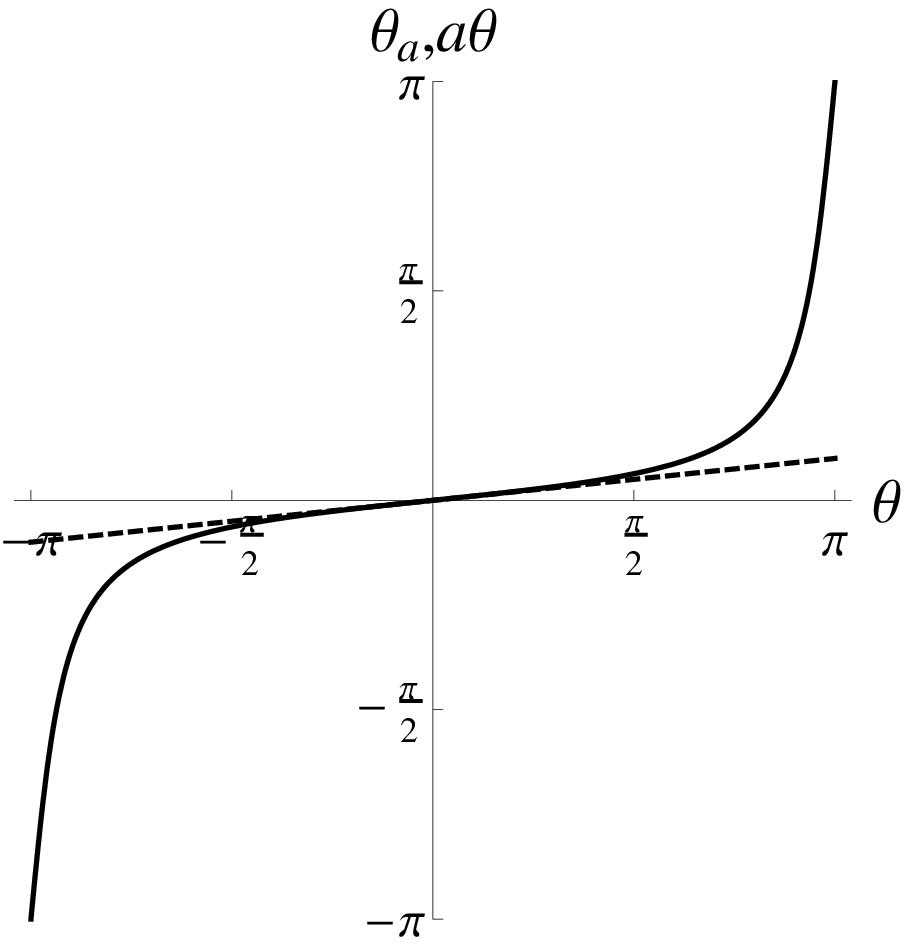}\hspace{5mm}
\includegraphics[height=4.5cm,width=5cm]{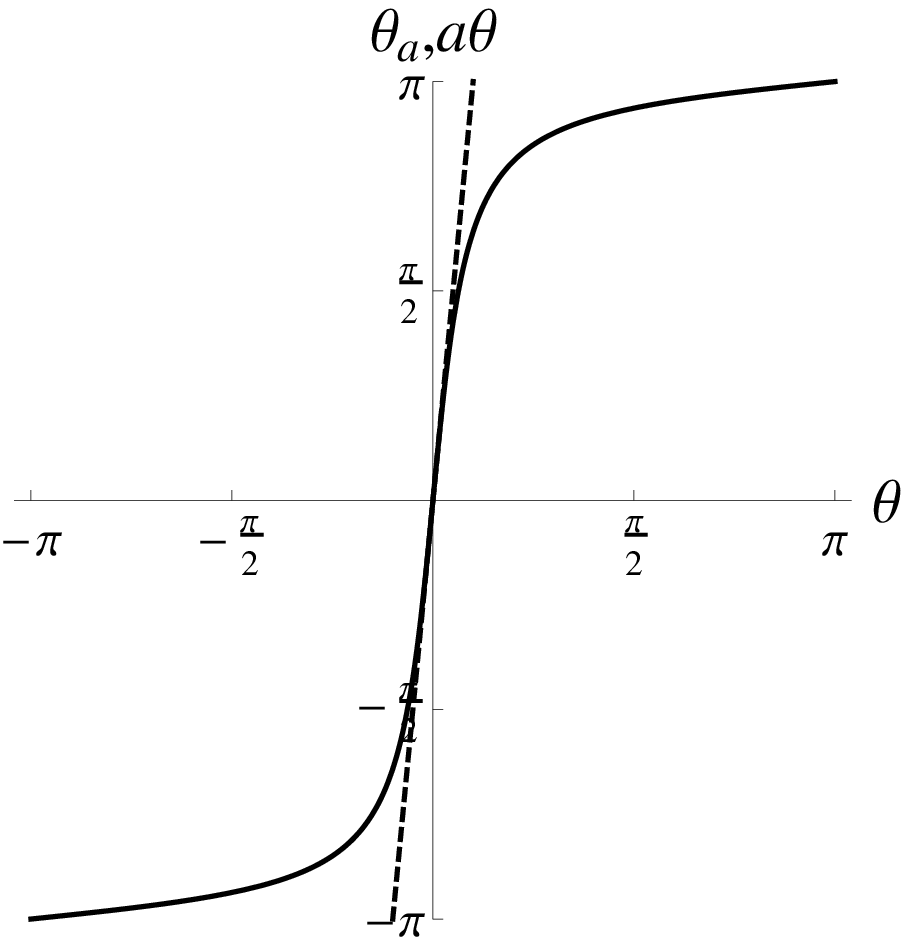}
\end{center}
\caption{From left to right: Illustration of the dilation given in \eqref{stereodiltoruseq} by stereographic projection.
Plot of $\theta_a$ (black) and $a \theta$ (dashed) as a function of $\theta$ for $a=0.1$ and $a=10$. Note that from the graphics it is
evident that $\theta_a$ and $\theta_{1/a}$ are inverse functions to each other.}\label{stereodiltorus}
\end{figure}
Note that this expression is similar to \eqref{dilS2angle} for the colatitude angle, but in our case $\theta_k\in(-\pi,\pi)$ instead of $(0,\pi)$.
As for the sphere, one can geometrically interpret this transformation as
independent dilations around the points $\theta_i=0\,,i=1,2$, lifted from the tangent lines to each (either meridian or equatorial) circle
by inverse stereographic projections (see Figure \ref{stereodiltorus}). 
For any function $f\in L^2(\mathbb T^2)$, a pure dilation  will be defined as

\begin{equation}
[D_{a_1,a_2}f](\theta_1,\theta_2)=\lambda(a_1,\theta_1)^{1/2}\lambda(a_2,\theta_2)^{1/2}
f((\theta_{1})_{1/a_1},(\theta_{2})_{1/a_2}),\label{funcdil}
\end{equation}
where
\begin{equation}
 \lambda(a,\theta)=\frac{d\theta_{1/a}}{d\theta}=\frac{2a}{(a^2-1)\cos\theta+a^2+1}\label{multitorus}
\end{equation}
is the Radon-Nikodym derivative, which is introduced to make the transformation \eqref{funcdil} unitary\footnote{Note that we are 
keeping the same symbol as 
for the multiplier of the sphere \eqref{multisphere}, even though they are different, since their respective measures are
different.}. In order to define wavelets, we also
incorporate translations with parameters $\vartheta_1,\vartheta_2\in (-\pi,\pi)$. Given $f\in  L^2(\mathbb T^2)$, one can prove that the action
$$f^{\vartheta_1,\vartheta_2}_{a_1,a_2}(\theta_1,\theta_2):= [U_{\vartheta_1,\vartheta_2}\circ D_{a_1,a_2} f](\theta_1,\theta_2),$$ explicitly written as
\begin{equation}
 {f}_{a_1,a_2}^{\vartheta_1,\vartheta_2}(\theta_1,\theta_2)=
 \lambda(a_1,\theta_1-\vartheta_1)^{1/2}\lambda(a_2,\theta_2-\vartheta_2)^{1/2}
 { f}((\theta_1-\vartheta_1)_{1/a_1},(\theta_2-\vartheta_2)_{1/a_2})\,,\label{reptorus}
\end{equation}
is unitary, where $D_{a_1,a_2}$ is given in \eqref{funcdil} and $U_{\vartheta_1,\vartheta_2}$ is the 
representation of translations on the torus.

As in the case of the sphere, we can characterize  admissible functions on the torus as follows:

\begin{definition}
A non-zero function $\gamma\in L^2(\mathbb T^2)$ is called admissible iff the condition
\begin{equation}
0<\int_X d\nu(\vartheta_1,\vartheta_2,a_1,a_2)|\langle {\gamma}_{a_1,a_2}^{\vartheta_1,\vartheta_2} |\psi\rangle|^2<\infty \label{admigeneralT2}
\end{equation}
is satisfied for any non-zero  $\psi \in L^2(\mathbb T^2)$, where
the measure on $X$ is
\begin{equation}
 d\nu(\vartheta_1,\vartheta_2,a_1,a_2)=\frac{da_1}{a_1^2}\frac{da_2}{a_2^2} \frac{d\vartheta_1 d\vartheta_2}{(2\pi)^2}.
\end{equation}
\end{definition}

The admissibility condition can be restated as follows:

\begin{proposition}\label{admiprop1}
  A non-zero function $\gamma\in L^2(\mathbb T^2)$ is admissible iff there exist $C\in\mathbb R$ such that
 \begin{equation}
0< \Lambda_{n_1,n_2}\equiv \int_0^\infty\int_0^\infty \frac{da_1}{a_1^2}\frac{da_2}{a_2^2}
|\widehat\gamma_{a_1,a_2}^{n_1,n_2}|^2<C< \infty \label{admistrongT2}
 \end{equation}
for all $(n_1,n_2)\in \mathbb{Z}^2$, where $\widehat\gamma_{a_1,a_2}^{n_1,n_2}=\langle \phi_{n_1,n_2}|\gamma_{a_1,a_2}\rangle$ are the Fourier coefficients of
$\gamma_{a_1,a_2}=D_{a_1,a_2}\gamma$.
\end{proposition}
\noindent{\bf Proof:}
The integral in the general admissibility condition \eqref{admigeneralT2} can be written as
\begin{eqnarray}
\int_0^\infty \int_0^\infty\frac{da_1}{a_1^2}\frac{da_2}{a_2^2}
\int_{-\pi}^{\pi} \int_{-\pi}^{\pi}\frac{d\vartheta_1d\vartheta_2}{(2\pi)^2} |\langle
\gamma^{\vartheta_1,\vartheta_2}_{a_1,a_2}|\psi\rangle|^2&=& \nonumber\\
\int_0^\infty\int_0^\infty\frac{da_1}{a_1^2}\frac{da_2}{a_2^2}\sum_{n_1,n_2=-\infty}^\infty
|\widehat{\gamma}^{n_1,n_2}_{a_1,a_2}|^2|\widehat{\psi}^{n_1,n_2}|^2  &=& \sum_{n_1,n_2=-\infty}^\infty
\Lambda_{n_1,n_2}|\widehat{\psi}^{n_1,n_2}|^2, \label{admiT2}
\end{eqnarray}
where we have used that $\langle \phi_{n_1,n_2}|\gamma^{\vartheta_1,\vartheta_2}_{a_1,a_2}\rangle=
e^{- i(n_1\vartheta_1 + n_2\vartheta_2)}\widehat{\gamma}^{n_1,n_2}_{a_1,a_2}$ and the usual orthogonality relations
for trigonometric functions, together with
 the definition (\ref{admistrongT2}) of $\Lambda_{n_1,n_2}$.

Taking into account
that $\{|\widehat{\psi}^{n_1,n_2}|^2\}\in \ell^1(\mathbb Z^2)$, since $\psi\in L^2(\mathbb{T}^2)$, the
admissibility condition (\ref{admigeneralT2}) adopts the following form:
\begin{equation}
 0<\sum_{n_1,n_2=-\infty}^\infty  |\widehat{\psi}^{n_1,n_2}|^2\Lambda_{n_1,n_2}<\infty,\;\;\forall \{|\widehat{\psi}^{n_1,n_2}|^2\}
\in \ell^1(\mathbb Z^2),\, \psi\not=0, 
\end{equation}
which converges absolutely iff $\{\Lambda_{n_1,n_2}\}\in \ell^\infty(\mathbb Z^2)$, that
is, iff $\Lambda_{n_1,n_2}<C<\infty$,  with $C$ independent of $n_1,n_2$. For the left inequality, it is required that $\Lambda_{n_1,n_2}>0$, which
proves the proposition.$\blacksquare$

This condition is not easy to verify. A simpler, but only necessary, condition is the following:

\begin{proposition}\label{propGamma}
A non-zero function $\gamma\in L^2(\mathbb T^2)$ is admissible only if it fulfills the condition
\begin{equation}
 \int_{-\pi}^\pi\int_{-\pi}^\pi \Gamma(\theta_1,\theta_2)
 d\theta_1d\theta_2=0\,.\label{admiweak}
\end{equation}
where
$\Gamma(\theta_1,\theta_2):=\gamma(\theta_1,\theta_2)/\sqrt{(1+\cos\theta_1)(1+\cos\theta_2)}$.
\end{proposition}
\noindent {\bf Proof:} Firstly, let us rewrite the expression of the Fourier coefficients
\begin{eqnarray}
 \widehat\gamma^{n_1,n_2}_{a_1,a_2}&=&\frac{1}{2\pi}\int_{-\pi}^{\pi} \int_{-\pi}^{\pi}d\theta_1d\theta_2 \gamma_{a_1,a_2}(\theta_1,\theta_2)
e^{-i(n_1\theta_1+n_2\theta_2)}\nonumber\\ &=& \frac{1}{2\pi}
\int_{-\pi}^{\pi} \int_{-\pi}^{\pi}d\theta_1d\theta_2 \lambda(a_1,\theta_1)^{1/2} \lambda(a_2,\theta_2)^{1/2}
\gamma(\theta_{1,1/a_1},\theta_{2,1/a_2})
e^{-i(n_1\theta_1+n_2\theta_2)}
\end{eqnarray}
by making the change of variables $\theta'_i=\theta_{i,1/a_i}$, and taking into account the multiplier property of the
Radon-Nikodym derivative
$\lambda(a,\theta_{1/a})^{-1}=\lambda(1/a,\theta)$, which results in
\begin{equation}
 \widehat\gamma^{n_1,n_2}_{a_1,a_2}=\frac{1}{2\pi}
\int_{-\pi}^{\pi} \int_{-\pi}^{\pi}d\theta_1d\theta_2 \lambda(1/a_1,\theta_1)^{1/2} \lambda(1/a_2,\theta_2)^{1/2}
\gamma(\theta_{1},\theta_{2})
e^{-i(n_1\theta_{1,a_1}+n_2\theta_{2,a_2})}.\label{gammahat}
\end{equation}
Actually, this change of variables has to do with the fact that $\widehat\gamma^{n_1,n_2}_{a_1,a_2}=
\langle \phi_{n_1,n_2}|D_{a_1,a_2}\gamma\rangle=\langle D_{1/a_1,1/a_2}\phi_{n_1,n_2}|\gamma\rangle$, that is, $D_{a_1,a_2}$ is unitary.

Let us evaluate the integral \eqref{admistrongT2} by splitting it into three regions: small, intermediate and large scales. 
For $a_i\ll 1$ we
can approximate
$\lambda(1/a_i,\theta_i)^{1/2}\approx\sqrt{2a_i}/\sqrt{1+\cos\theta_i}$.
Let us assume that the support $S_\gamma$ of $\gamma$ does not
contain $(\pm \pi,\pm \pi)$,
so that $\lim_{a_i\to 0}\theta_{i,a_i}=0, \forall\theta_i\in S_\gamma$ and we have 
$e^{-i(n_1\theta_{1,a_1}+n_2\theta_{2,a_2})}\to 1\,\, \forall n_1,n_2\in\mathbb Z$  in this limit. 
Thus, the integral
\eqref{admistrongT2} over small
scales $0<a_i<\epsilon_i\ll 1$ can be written as
\begin{equation}
 \int_0^{\epsilon_1} \int_0^{\epsilon_2}\frac{da_1}{a_1}\frac{da_2}{a_2}\left|\int_{-\pi}^{\pi} \int_{-\pi}^{\pi}d\theta_1d\theta_2 \frac{\gamma(\theta_1,\theta_2)}
{\sqrt{1+\cos\theta_1}\sqrt{1+\cos\theta_2}}\right|^2<\infty,
\end{equation}
which implies \eqref{admiweak}. 

For intermediate scales, since $D_{a_1,a_2}$ is a strongly continuous operator
and by the continuity of the scalar product, we have that the integrand in \eqref{admistrongT2} is a bounded continuous
function in this region. 

For large scales, from \eqref{gammahat} we can bound
\begin{equation}
 |\widehat\gamma^{n_1,n_2}_{a_1,a_2}|\leq\frac{\mathrm{sup}(|\gamma|)}{2\pi}
\int_{-\pi}^{\pi} \int_{-\pi}^{\pi}d\theta_1d\theta_2 \lambda(1/a_1,\theta_1)^{1/2} \lambda(1/a_2,\theta_2)^{1/2},\label{gammahat2}
\end{equation}
where $\mathrm{sup}(|\gamma|)$ denotes the supremum of $|\gamma|$. The integral 
\begin{equation}
 \int_{-\pi}^{\pi}d\theta\;\lambda(1/a,\theta)^{1/2}=\frac{4 K(1-\frac{1}{a^2})}{\sqrt{a}}
\end{equation}
is written in terms of the complete elliptic integral of the first kind $K$, whose large scale behavior is given by
\begin{equation}
K\left(1-\frac{1}{a^2}\right)\sim \ln(a), \;\; a\gg 1,
\end{equation}
so that the integral \eqref{admistrongT2} over large scales converges as well.

Finally, if we drop the restriction on the support of $\gamma$, the condition
\eqref{admiweak} is only necessary, which proves the proposition.$\blacksquare$

In general, an admissibility condition does not guarantee a proper reconstruction of a function from its
wavelet coefficients, and a frame condition is required. However, as in the standard case, the admissibility
condition \eqref{admistrongT2}
is enough. We shall consider  localized admissible functions $\gamma$ in order to provide an easier proof.
By ``localized'' we mean that $\theta_{i,a_i}\approx a_i\theta_i, \forall (\theta_1,\theta_2)\in S_\gamma$ and $a_i\leq 1$
(i.e., a valid approximation in the Euclidean limit). For practical purposes, this is not really a restriction since 
the approximation  $\theta_a\approx a\theta$ is quite good for a large range of $\theta$ when $a\leq 1$, see Figure \ref{stereodiltorus}.

Let us denote by $Q_q, q=1,2,3,4$, the four quadrants 
of the  Fourier plane in counterclockwise order. Since dilations do not mix quadrants, and translations do not 
change the support of $\widehat\gamma$, it is clear that $\widehat\gamma$ 
must have support on all (four)  
quadrants in order to be admissible. Under these assumptions, one has the following result:

\begin{theorem}\label{continuousf} For any localized admissible function $\gamma$, the family
$\{\gamma_{a_1,a_2}^{\vartheta_1,\vartheta_2},\ (\vartheta_1,\vartheta_2,a_1,a_2)\in X\}$ is a continuous
frame; that is, there exist real constants $0<c\leq C$ such that
\begin{equation}
c||\psi||^2\leq \int_X  d\nu(\vartheta_1,\vartheta_2,a_1,a_2)
|\langle \gamma_{a_1,a_2}^{\vartheta_1,\vartheta_2}|\psi\rangle|^2\leq C ||\psi||^2,\;\;\forall
\psi\in L^2(\mathbb T^2) .\label{lowerupperb}
\end{equation}
\end{theorem}
\noindent{\bf Proof:} It remains only to
prove the lower bound, which is equivalent to prove that the quantity defined in
(\ref{admistrongT2}) is uniformly bounded from below: $\Lambda_{n_1,n_2}>c,\,\forall n_1,n_2\in\mathbb{Z}$.

Since $\gamma_{a_1,a_2}$ are integrable on $\mathbb{T}^2$, their Fourier coefficients $\widehat{\gamma}_{a_1,a_2}^{n_1,n_2}$ 
tend to zero for $|n_1|,|n_2|\to\infty$,
which implies that the problematic region is now that for which  $|n_1|,|n_2|\gg 1$. 
Let us focus on the $a\ll 1$ region. Using that $\gamma$ is localized, we can write
$\lambda(1/a_i,\theta_i)^{1/2}=\sqrt{2a_i}/\sqrt{1+\cos\theta_i} +O(a_i^{5/2})$, where the error term is bounded, and
$\theta_a\approx a\theta$ for small $a$. Within this approximation, the expression 
\eqref{gammahat} reads 
\begin{equation}
\widehat{\gamma}_{a_1,a_2}^{n_1,n_2}\approx 2 \sqrt{a_1a_2} \,\widehat{\Gamma}^{a_1n_1,a_2n_2},\label{smallscales}
\end{equation}
where $\Gamma$ is introduced in Proposition \ref{propGamma}, this estimation being valid as long as 
$\widehat{\Gamma}^{a_1n_1,a_2n_2}\neq 0$ (which is the interesting case for us). Note that when writing $\widehat{\Gamma}^{a_1n_1,a_2n_2}\equiv\widehat{\Gamma}^{\alpha_1,\alpha_2}$ we are extending the 
integer Fourier indices to the reals $\alpha_1,\alpha_2\in\mathbb R$ in a 
continuous (and differentiable) way as a consequence of Lebesgue's dominated convergence theorem.  
For $q=1,\ldots,4$, let $(\alpha_1^0,\alpha_2^0)\in Q_q$ such that $|\widehat\Gamma^{\alpha_1^0,\alpha_2^0}|>0$, in particular, 
we can chose the values of $\alpha_1^0,\alpha_2^0$ where the maximum of $|\widehat\Gamma^{\alpha_1,\alpha_2}|$ in the current quadrant $Q_q$ 
is attained. Since $|\widehat\Gamma^{\alpha_1,\alpha_2}|$ is continuous  there exist $\rho_i$, with $0<\rho_i<|\alpha_i^0|,\,i=1,2$,  
such that $|\widehat\Gamma^{\alpha_1,\alpha_2}|> |\widehat\Gamma^{\alpha_1^0,\alpha_2^0}|/2$  in the region 
$R=(\alpha_1^0-\rho_1,\alpha_1^0+\rho_1)\times (\alpha_2^0-\rho_2,\alpha_2^0+\rho_2)\subset Q_q$. Considering $|n_1|,|n_2|\gg 1$, we have that 
\begin{eqnarray}
 \Lambda_{n_1,n_2}&=&\int_0^\infty\int_0^\infty \frac{da_1}{a_1^2}\frac{da_2}{a_2^2}
|\widehat\gamma_{a_1,a_2}^{n_1,n_2}|^2 \geq  \int_{\alpha_1^0-\rho_1}^{\alpha_1^0+\rho_1} 
\int_{\alpha_2^0-\rho_2}^{\alpha_2^0+\rho_2}
\frac{d\alpha_1}{\alpha_1}\frac{d\alpha_2}{\alpha_2}   
4|\widehat\Gamma^{\alpha_1,\alpha_2}|^2 \nn\\
&>& |\widehat\Gamma^{\alpha_1^0,\alpha_2^0}|^2 \log\frac{\alpha_1^0+\rho_1}{\alpha_1^0-\rho_1}
\log\frac{\alpha_2^0+\rho_2}{\alpha_2^0-\rho_2}\,,\label{framebound}
\end{eqnarray}
Note that $\alpha_1^0,\alpha_2^0$ being fixed, and  $|n_1|,|n_2|\gg 1$, 
gives $a_1=\alpha_1/|n_1|, a_2=\alpha_2/|n_2|$ small for $\alpha_1,\alpha_2\in R$, which justifies the approximation 
\eqref{smallscales}.  Thus \eqref{framebound} gives a strictly positive quantity independent of
$n_1,n_2$ in each quadrant, which proves that
$\Lambda_{n_1,n_2}$ is bounded from below. $\blacksquare$

%

The CWT of a function $\psi\in L^2(\mathbb T^2)$ reads as:

\begin{equation}
 \Psi^{\vartheta_1,\vartheta_2}_{a_1,a_2}=\langle \gamma^{\vartheta_1,\vartheta_2}_{a_1,a_2}|\psi\rangle=\iint_{\mathbb T^2} 
 \overline{\gamma^{\vartheta_1,\vartheta_2}_{a_1,a_2}(\theta_1,\theta_2)}\psi(\theta_1,\theta_2)d\omega,\;
\psi\in L^2(\mathbb T^2).\label{wavecoef}
\end{equation}

The original function $\psi$ can be reconstructed (in the weak sense) from its wavelet coefficients $\Psi^{\vartheta_1,\vartheta_2}_{a_1,a_2}$ 
by means of the reconstruction formula:
\begin{equation} 
\psi(\theta_1,\theta_2)=\int_X d\nu(a_1,a_2,\vartheta_1,\vartheta_2)  
\Psi^{\vartheta_1,\vartheta_2}_{a_1,a_2} \;\widetilde\gamma^{\vartheta_1,\vartheta_2}_{a_1,a_2}(\theta_1,\theta_2)
\label{reconstruction}
\end{equation}
where $\{\widetilde\gamma^{\vartheta_1,\vartheta_2}_{a_1,a_2}\}$ is the dual frame (see e.g. chapter 5 of \cite{Ole} for the general definition) whose Fourier coefficients 
are given by
\begin{equation}
 \langle \phi_{n_1n_2}|\widetilde\gamma^{\vartheta_1,\vartheta_2}_{a_1,a_2}\rangle= \Lambda^{-1}_{n_1n_2}
\langle \phi_{n_1n_2}|\gamma^{\vartheta_1,\vartheta_2}_{a_1,a_2}\rangle.
\end{equation}

Note that the dual frame is well-defined ($0\neq\widetilde\gamma^{\vartheta_1,\vartheta_2}_{a_1,a_2}\in L^2(\mathbb{T}^2)$) 
since Theorem \ref{continuousf} ensures that  $0<c<\Lambda_{n_1n_2}<C<\infty,\,\forall (n_1,n_2)\in \mathbb{Z}^2$.


\subsection{Existence of admissible functions}

Now we discuss the existence of admissible functions on the torus fulfilling \eqref{admistrongT2}. For this purpose, we
shall resort to Euclidean wavelets. Wavelets on the plane $\mathbb R^2$ with two dilations can be defined through the
natural tensor product representation (see e.g. chapter 5 of \cite{tensor}), where a unitary representation of the affine group in $L^2(\mathbb R^2)\ni\psi$ is given by
\begin{equation}
\psi_{a_1,a_2}^{b_1,b_2}=[U(a_1,a_2,b_1,b_2) \psi](x_1,x_2)=a_1^{-1/2}a_2^{-1/2} 
\psi\left(\frac{x_1-b_1}{a_1},\frac{x_2-b_2}{a_2}\right).\label{tpaffine}
\end{equation}
The ``tensor-product'' admissibility condition for $\psi\in L^2(\mathbb R^2)$  adopts the following form
\begin{equation}
 \int_{-\infty}^\infty\int_{-\infty}^\infty \frac{\widehat{\psi}(k_1,k_2)}{|k_1|\,|k_2|}dk_1 dk_2<\infty,\label{tpadR2}
\end{equation}
where by $\widehat{\psi}$ we mean the Fourier transform of $\psi$. It can be easily checked that if $\psi_1(x_1), \psi_2(x_2)\in L^2(\mathbb R)$ are
admissible functions generating standard wavelet frames, with frame bounds $c_i, C_i, i=1,2$, then the tensor product
$\psi(x_1,x_2)=\psi_1(x_1)\psi_2(x_2)$ fulfills \eqref{tpadR2} and generates a tensor wavelet  frame (under the group action \eqref{tpaffine})
in $L^2(\mathbb R^2)$, with frame bounds $c_1c_2$ and $C_1C_2$. Note that $\psi(x_1,x_2)$ does not necessarily need to be 
a product of the form $\psi_1(x_1)\psi_2(x_2)$, although functions of this kind span  $L^2(\mathbb R^2)$. 
\begin{proposition}
 A ``tensor-product'' admissible function $\psi\in L^2(\mathbb R^2)$ provides an admissible function on
$L^2(\mathbb T^2)$, fulfilling \eqref{admistrongT2},  by
inverse stereographic projection
\begin{equation}
 [\Pi^{-1}_{\mathbb T^2}\psi](\theta_1,\theta_2)=\frac{1}{\sqrt{1+\cos\theta_1}\sqrt{1+\cos\theta_2}}
\psi\left(2\tan\frac{\theta_1}{2},2\tan\frac{\theta_2}{2}\right). \label{inversestereo}
\end{equation}
\end{proposition}
The proof is direct.

\begin{figure}
\begin{center}
\includegraphics[height=4cm]{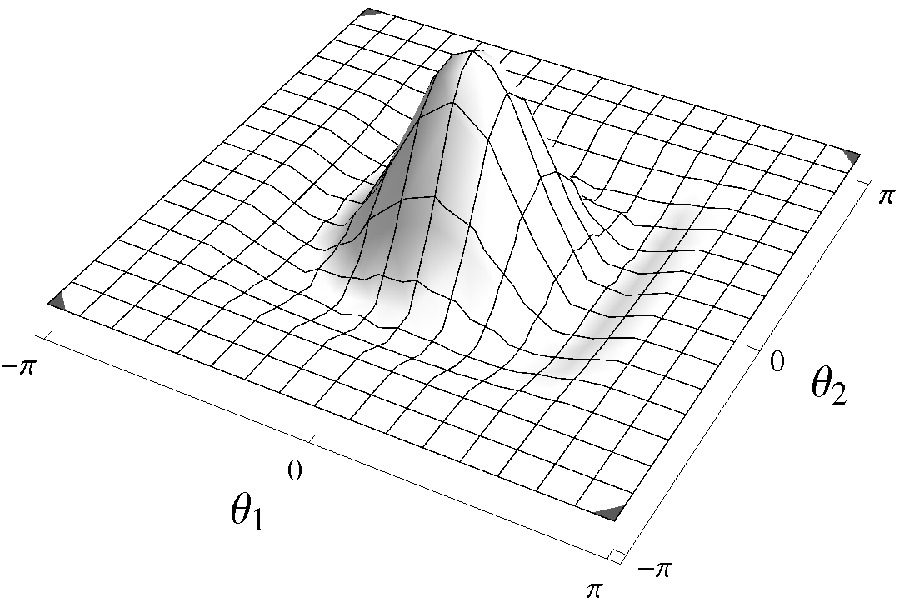}
\includegraphics[height=4cm]{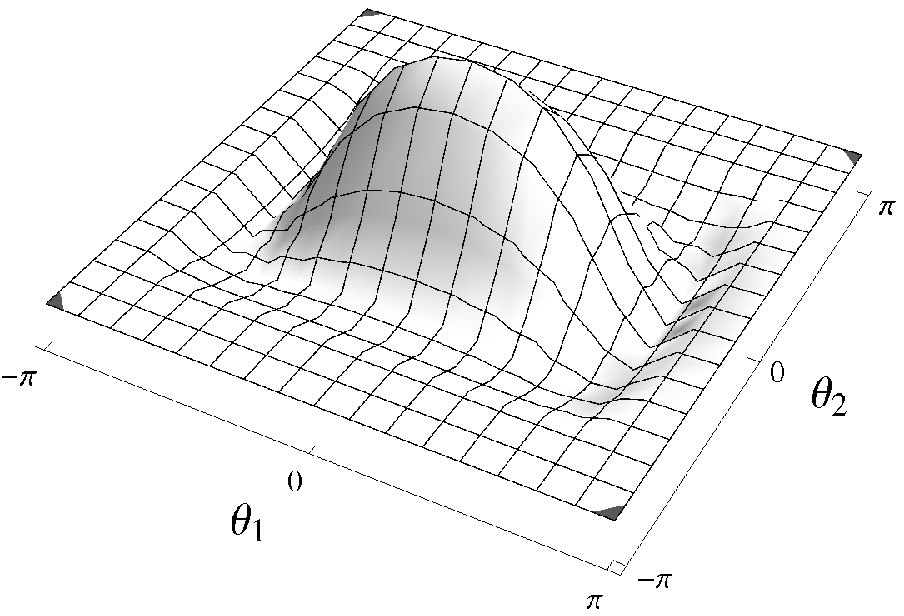}\includegraphics[height=4cm]{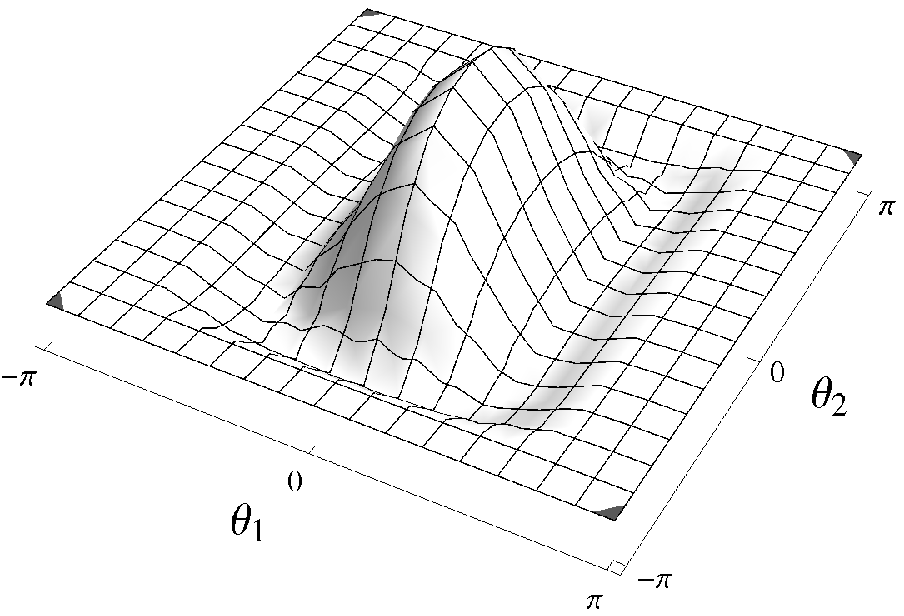}
\end{center}
\caption{Dilation of the axisymmetric DOG on $\mathbb T^2$ with $\alpha=2$ for (from left to right): $a_1=a_2=1$; $a_1=2, a_2=1$ and 
$a_1=1, a_2=2$. }\label{DOGaxidil}
\end{figure}

Let us provide some explicit examples of admissible functions on  $L^2(\mathbb T^2)$ imported from  $L^2(\mathbb R^2)$ by 
inverse sthereographic projection. For this purpose we shall consider Difference of Gaussians (DoG), commonly 
used as a pass-band filter in image science, which in one dimension are written as
\begin{equation}
 \psi_\alpha(x)=e^{-x^2}-\frac{e^{-x^2/\alpha^2}}{\alpha}.\label{DoG}
\end{equation}
For a two-dimensional {\it separable} DoG function $\psi_{\alpha_1,\alpha_2}(x_1,x_2)=\psi_{\alpha_1}(x_1)\psi_{\alpha_2}(x_2)$, the inverse sthereographic 
projection  \eqref{inversestereo} leads to the function
\begin{equation}
[\Pi^{-1}_{\mathbb T^2}\psi_{\alpha_1,\alpha_2}](\theta_1,\theta_2)=\frac{1}{\sqrt{1+\cos\theta_1}\sqrt{1+\cos\theta_2}}
\psi_{\alpha_1}\left(2\tan\frac{\theta_1}{2}\right)\psi_{\alpha_2}\left(2\tan\frac{\theta_2}{2}\right).
\end{equation}
Usually the axisymmetric (non-separable) DoG
\begin{equation}
\psi_\alpha(x_1,x_2)=e^{-(x_1^2+x_2^2)}-\frac{e^{-(x_1^2+x_2^2)/\alpha^2}}{\alpha^2}.
\end{equation}
is considered in two dimensions. For this case, the corresponding function on $\mathbb T^2$ is explicitly 
 \begin{equation}
[\Pi^{-1}_{\mathbb T^2}\psi_\alpha](\theta_1,\theta_2)= \frac{1}{\sqrt{1+\cos\theta_1}\sqrt{1+\cos\theta_2}}
\psi_{\alpha}\left(2\tan\frac{\theta_1}{2},2\tan\frac{\theta_2}{2}\right).\label{axiDoGT2}
 \end{equation}
In Figure \ref{DOGaxidil} we represent the axisymmetric DoG on $\mathbb T^2$ \eqref{axiDoGT2} and its dilation  
\eqref{funcdil} for  two cases: $a_1=2, a_2=1$ and 
$a_1=1, a_2=2$, respectively.

One would expect the wavelet transform on the torus to behave locally
(at short scales or large values of the equatorial and longitudinal radius $R_1,R_2\to\infty$)
like the standard wavelet transform on the plane. In fact, in the Euclidean limit $R_1,R_2\to\infty$, 
which is given by two copies of the Euclidean limit on the circle \cite{wavecircle}, one recovers the 
tensor product wavelet construction 
on the plane (\ref{tpadR2}, \ref{tpaffine}).

Note that, since rotations are absent in the torus, when proving Theorem \ref{continuousf} it has been essential 
to have two dilations $a_1, a_2$ at our disposal. 
Indeed, we need two different dilations to bring any pair $(n_1,n_2)$ to the small rectangle $R$ where the extension of 
$\widehat{\Gamma}$ to the reals is non-zero, thus ensuring that $\Lambda_{n_1,n_2}>c$ in \eqref{framebound}. 

However, wavelet constructions on the plane with a single dilation are customary (see for example curvelets
\cite{curveletrefs}  shearlets \cite{shearletrefs}, etc). Actually, one could restrict himself 
to a ``single'' dilation $(a_1,a_2=\sigma(a_1))$, with $\sigma$ a strictly positive increasing function, usually $\sigma(a)=a$, although 
other choices like, for example, ``parabolic'' dilations $\sigma(a)=\sqrt{a}$ are used for shearlets. This implies a 
restriction of  the parameter space $X$ to
$X'=\{(a,b_1,b_2), a>0, b_{1,2}\in \mathbb R\}$. From the measure $d
\nu(b_1,b_2,a_1,a_2)=db_1db_2\frac{da_1^2}{a_1^2}\frac{da_2^2}{a_2^2}$ on $X$ we derive the measure on $X'$
\begin{equation}
 d\nu'(b_1,b_2,a)=\sigma(a)\frac{da}{a^4}
{db_1 db_2}.
\end{equation}
%

 The problem now is whether the  subset 
$\{ \psi_{a}^{b_1,b_2}\equiv\psi_{a,\sigma(a)}^{b_1,b_2}\}$ in \eqref{tpaffine} is a frame or not. The proof of frame condition 
for the plane is similar to the proof of frame condition for the torus given in Theorem \ref{continuousf}, 
with obvious modifications ($\theta_{1,2}\to b_{1,2}, n_{1,2}\to k_{1,2}$ and $\widehat\Gamma\rightarrow \widehat\psi$, etc.). 
As already said,  we need two different dilations to bring any pair $(k_1,k_2)$ to the small rectangle $R$ where 
$\widehat{\psi}$ is non-zero, thus ensuring that $\Lambda_{k_1,k_2}>c$ like in \eqref{framebound}. A way out could be to impose 
additional conditions to the support of $\widehat\psi$, like  extending it to a ring around the origin 
$(0,0)$ \cite{waveletmicro}, or to introduce extra group parameters like rotations, shears, etc. Also, in the discrete case, 
frames in  $\mathbb R^n$, with $n\geq 2$, with a single dilation are constructed from more 
than one (in fact at least $2^{n}-1$) admissible function \cite{Gustavo,weiss}.

%


\section{Modular wavelets}\label{secmodular}

In this section we shall pursue the use of the modular group as an extra set of wavelet parameters on the torus. This option has the 
advantage that we do not need to enlarge the support of $\widehat\Gamma$ but, on the contrary, it can be restricted to 
a one-dimensional subset. Actually, when modular transformations are introduced, a frame condition can be 
proved when  setting $\sigma(a)=a$ and considering the case  $\widehat{\Gamma}^{n_1,n_2}=0, \forall n_1\not=n_2
\in\mathbb Z$, which means that $\Gamma(\theta_1,\theta_2)=
\eta(\theta_1+\theta_2)$ for some function $\eta:\mathbb S^1\to \mathbb C$, although other choices are also possible like 
$\Gamma(\theta_1,\theta_2)=\eta(\theta_1)$ or $\Gamma(\theta_1,\theta_2)=\eta(\theta_2)$. 

Before entering into the discussion of ``modular wavelets'', we shall make a small introduction to modular transformations and 
modular frames.
\subsection{Modular group on the Torus ${\mathbb T}^2$}

In this subsection we introduce the modular group on the torus and give its main properties.

\begin{definition}
 The modular group on the torus ${\mathbb T}^2$
is the subgroup
\begin{equation}
SL(2,\mathbb Z)=\left\{M=\begin{pmatrix} m & n\\ p& q\end{pmatrix};
 m,n,p,q\in\mathbb Z, \;\det(M)=mq-np=1\right\},
\end{equation}
of the group  $SL(2,\mathbb{R})$ of linear transformations of the plane preserving the area 
with integer entries.
\end{definition}

The modular group
transforms pair of integers $(n_1,n_2)$ into
pairs of integers $(n'_1,n'_2)^t=M(n_1,n_2)^t=(m n_1+n n_2,p n_1 + q n_2)^t$. Therefore it
preserves the torus
${\mathbb T}^2=\mathbb R^2/\mathbb Z^2$, and  its action can be lifted to functions on the torus
in the ordinary way:
\begin{equation}
 f_M(\theta_1,\theta_2)\equiv f(M^{-1}(\theta_1,\theta_2)^t).
\end{equation}
Since $M$ preserves the area, this defines a unitary representation of
$SL(2,\mathbb Z)$ on $L^2(\mathbb T^2)$:
\begin{eqnarray}
 U: & & L^2(\mathbb T^2)\rightarrow L^2(\mathbb T^2) \nonumber\\
 & & f(\theta_1,\theta_2)\mapsto [U(M)f](\theta_1,\theta_2)\equiv f_M(\theta_1,\theta_2)\,.
\label{repremodular}
\end{eqnarray}

However, this unitary representation is not irreducible, admitting infinite invariant subspaces
${\cal V}_g\subset L^2(\mathbb{T}^2), g\in\mathbb{N}\cup\{0\}$. To prove this, we first 
state the following Lemma, whose proof is immediate using that modular
transformations are area preserving:

\begin{lema}
 The action of the modular group in Fourier space is given by:
\begin{equation}
  \widehat f_M^{(n_1,n_2)}= \widehat f^{(n_1,n_2)M}\quad
\forall (n_1,n_2)\in \mathbb{Z}^2\,,\,M\in SL(2,\mathbb Z)\,,\,f\in L^2(\mathbb{T}^2)\,.
\end{equation}
\label{lemma-modularfourier}\end{lema}
This means that the action of a modular transformation $M$ in Fourier space is through its transpose
$\vec{n}'=M^{t}\vec{n}$, which is again a modular transformation. Since we shall work mainly in Fourier space, and to simplify
 notation, we shall consider the action on row vectors, $(n'_1,n'_2)=(n_1,n_2)M$. To obtain the corresponding action
for column vectors, a transpose operation should be performed.

The action of the
modular group on $\mathbb{Z}^2$ is not transitive, leaving certain  subsets invariant, as stated in
the following  Lemma, also easy to prove. In what follows, g.c.d. stands for greatest common divisor.

\begin{lema}
 The subsets ${\cal G}_g=\{(n_1,n_2)\in \mathbb{Z}^2\,:\, \hbox{\rm g.c.d.}(n_1,n_2)=g\}$,
with ${\cal G}_0\equiv \{(0,0)\}$, are invariant under the modular group. 
\label{lemma-modularinvariant}
\end{lema}

With the aid of Lemma \ref{lemma-modularfourier} and Lemma \ref{lemma-modularinvariant}, the
following proposition is easy to prove:

\begin{proposition}
The subspaces ${\cal V}_g\,,\, g=0,1,2,\ldots$ of  $L^2(\mathbb{T}^2)$ given by
\begin{equation}
 {\cal V}_g = \{\psi\in L^2(\mathbb{T}^2) \,:\, \mathrm{supp}(\widehat{\psi})\subset  {\cal G}_g\}
\end{equation}
are invariant under the action of the modular group $SL(2,\mathbb{Z})$.
\end{proposition}

We can think of $\mathbb{Z}^2$ as partitioned into orbits under the action of
$SL(2,\mathbb{Z})$. Each  orbit ${\cal G}_g$ is generated by the action of the group on, let us 
say, the point $(g,g)\in \mathbb{Z}^2$. The action of the modular group in each orbit ${\cal G}_g$
is transitive but not free, since the point $(g,g)\neq (0,0)$ has a stabilizer (or isotropy) group 
that is given by:
\begin{equation}
 N=\left\{\begin{pmatrix} 2 & 1\\ -1& 0\end{pmatrix}^k,\; k\in\mathbb Z\right\}\sim \mathbb Z\label{isotropydiag}
\end{equation}
while the point $(0,0)$, which is an orbit by itself, has as stabilizer the whole group $SL(2,\mathbb{Z})$.
Note that the stabilizer is the same for all orbits ${\cal G}_g\,,g\neq0$. Also, for $g\neq0$, if we
choose a different point in the orbit (like $(g,0)$ or $(0,g)$), the stabilizer group is 
different
but isomorphic (in fact conjugate). For example, for $(g,0)$, the stabilizer is
\begin{equation}
 N_1=\left\{\begin{pmatrix} 1 & 0\\ 1& 1\end{pmatrix}^k,\; k\in\mathbb Z\right\}\sim \mathbb Z,\label{isotropy1}
\end{equation}
while for $(0,g)$ it is
\begin{equation}
 N_2=\left\{\begin{pmatrix} 1 & 1\\ 0& 1\end{pmatrix}^k,\; k\in\mathbb Z\right\}\sim \mathbb Z.\label{isotropy2}
\end{equation}

By the orbit-stabilizer theorem (see e.g. chapter 10 of  \cite{orbit-stabilizer-theorem}),
there is a bijection between each orbit ${\cal G}_g\,,g\neq0$, and the quotient
${\cal X}\equiv SL(2,\mathbb{Z})/N$. This means that there is also a bijection between each pair
of orbits ${\cal G}_g,\,{\cal G}_{g'}$ with $g,g'\neq0$. This bijection can be realized as follows:


\begin{proposition}
Given $(n_1,n_2)\in {\cal G}_g$, there is only one representative $M^g_{n_1,n_2}\in \mathcal{X}$ (i.e. modulo N) such
that  $(n_1,n_2)M^g_{n_1,n_2}=(g,g)$.
\end{proposition}
\textbf{Proof:}
 We can pick the representative
\begin{equation}
 M^g_{n_1,n_2}=\begin{pmatrix} m &  m-n_2/g\\ n & n+n_1/g\end{pmatrix},\label{representative}
\end{equation}
where $m,n$ fulfill B\'ezout's identity $mn_1+nn_2=g$ and can be easily computed with the extended
Euclidean algorithm. All other elements $M\in SL(2,\mathbb{Z})$ transforming $(n_1,n_2)$ into
$(g,g)$ can be obtained  by
multiplying $M^g_{n_1,n_2}$ by elements in $N$.$\blacksquare$

It should be stressed that $M^g_{n_1,n_2}$ can be written as $M^g_{n_1,n_2}=M^1_{n'_1,n'_2}$, where $ n'_1=n_1/g\,,n'_2=n_2/g$ 
are coprime, i.e. g.c.d.$(n'_1,n'_2)=1$. This allows us to take the representative $M_{n'_1,n'_2}\equiv M^1_{n'_1,n'_2}=M^g_{n_1,n_2}$ for all cases $g\neq 0$,
for instance, when writing expressions like $\sum_{M\in{\cal X}}$.

Note that similar results hold for $(g,0)$ and $(0,g)$.

The previous proposition allows us to label pairs $(n_1,n_2)\in \mathbb{Z}^2$ equivalently as $(g,M_{\frac{n_1}{g},\frac{n_2}{g}}^{-1})$, where g.c.d$(n_1,n_2)=g$,  
for $(n_1,n_2)\neq (0,0)$; for  $(n_1,n_2)=(0,0)$ we can label it as $(g=0,I_2)$, where $I_2$ represents the $2\times 2$ identity matrix..

All this construction translates, \textit{mutatis mutandis}, to the subspaces ${\cal V}_g$,
that are orbits through, let us say, $\phi_{g,g}$ (defined in \eqref{ONBtorus}), by the action of
the modular group. The action of the modular group in each orbit is transitive but not free,
the stabilizer group being again $N$ for orbits ${\cal V}_g\,,g\neq0$, and the whole $SL(2,\mathbb{Z})$ for ${\cal V}_0$.
There is a bijection between each orbit ${\cal V}_g\,,g\neq0$ and the quotient
${\cal X}\equiv SL(2,\mathbb{Z})/N$, and between each pair
of orbits ${\cal V}_g,\,{\cal V}_{g'}$ with $g,g'\neq0$. Thus, expressions like 
$\sum_{n_1,n_2=-\infty}^{\infty}q_{n_1,n_2}$ can be written as $\sum_{g=0}^{\infty}\sum_{M\in{\cal X}_g}q_{g,M^{-1}}$, where we 
mean by $\mathcal{X}_0=\{I_2\}$ and $\mathcal{X}_g=\mathcal{X}$ for $g\neq 0$. 
We hope that this slight abuse of notation does not create confusion.

The previous considerations can be restated as follows:
\begin{proposition}
Let $g\in\mathbb{N}$.  If $\gamma=\phi_{n_1,n_2}$, with g.c.d$(n_1,n_2)=g$, then $B_{g,\gamma}=\{\gamma_M\,/\,M\in {\cal X}\}$   
is an orthonormal basis of ${\cal V}_g$.
\end{proposition}
\textbf{Proof:} This is a consequence of the unitarity and irreducibility of the representation $U$ of $SL(2,\mathbb{Z})$
in \eqref{repremodular} restricted to ${\cal V}_g$, and that we restrict the action to the quotient ${\cal X}$,
otherwise divergences would occur due to the ``infinite measure'' of the non-compact  subgroup
$N$. In the terminology of \cite{Gazeau}, the representation is square integrable modulo
$(\sigma,N)$, where $\sigma$ is a Borel section from ${\cal X}$ to $SL(2,\mathbb{Z})$.$\blacksquare$

The question is whether we can extend this ``basis'' to the whole $L^2(\mathbb{T}^2)$. The answer is given in the
following Proposition:

\begin{proposition}\label{propBessel}
 Let $\eta\in L^2(\mathbb{T}^1)$ such that {\rm supp}$(\widehat\eta)=\mathbb{Z}$, and define
$\gamma(\theta_1,\theta_2)=\eta(\theta_1+\theta_2)$. Then the set
$F_{\gamma}=\{\gamma_M^{\vartheta_1,\vartheta_2}\,/\,M\in {\cal X}\,,\,\vartheta_1,\vartheta_2\in\mathbb{T}^2\}$ is a 
complete Bessel sequence (see e.g. chapter 3 of \cite{Ole}) in $L^2(\mathbb{T}^2)$, in the sense
that there exist $C>0$ such that
\begin{equation}
0< \int\frac{d\vartheta_1d\vartheta_2}{(2\pi)^2}\sum_{M\in{\cal X}} |\langle \gamma_M^{\vartheta_1,\vartheta_2}
|\psi\rangle|^2 \leq  C ||\psi||^2\,,\,\,\forall \psi\in L^2(\mathbb{T}^2),\, \psi\neq 0\,.
\label{Bessel1}
\end{equation} \label{modularBessel}
\end{proposition}
\textbf{Proof:} Using the same steps as in Proposition \ref{admiprop1}, making use of the reparametrization 
of the sum $\sum_{n_1,n_2=-\infty}^{\infty}$ in terms of $g={\mathrm g.c.d.}(n_1,n_2)$ and $M'\in\mathcal{X}$ 
given before, denoting $\widehat\psi^{g,M'^{-1}}\equiv\widehat\psi^{n_1,n_2}$, and taking into account that 
$\widehat\gamma^{n_1,n_2}=\widehat\gamma^{g,g}\delta_{n_1,g}\delta_{n_2,g}$, we can write
%
%
\begin{eqnarray}
&& \int\frac{d\vartheta_1d\vartheta_2}{(2\pi)^2}\sum_{M\in{\cal X}} |\langle \gamma_M^{\vartheta_1,\vartheta_2}
|\psi\rangle|^2  = 
\sum_{g=0}^\infty \sum_{M'\in{\cal X}_g}  |\widehat\psi^{g,M'^{-1}}|^2  \sum_{M\in{\cal X}} 
|\widehat\gamma^{g,M'^{-1}}_M|^2= \nn\\ && \sum_{g=0}^\infty \sum_{M'\in{\cal X}_g}  |\widehat\psi^{g,M'^{-1}}|^2 |\widehat\gamma^{g,g}|^2 =
\sum_{g=0}^\infty |\widehat\gamma^{g,g}|^2 \|P_g\psi\|^2\leq \mathrm{max}_{g}\{|\widehat\gamma^{g,g}|^2\} \|\psi\|^2 \,,\label{Bessel2}
\end{eqnarray}
where we have used that the only term contributing to the sum 
\begin{equation}
\sum_{M\in{\cal X}}|\widehat\gamma_M^{g,M'^{-1}}|^2=
\sum_{M\in{\cal X}}|\widehat\gamma^{g,M'^{-1}M}|^2=|\widehat\gamma^{g,g}|^2
\end{equation}
is $M_{n_1,n_2}$. We have also used the Parseval identity $\sum_{M'\in{\cal X}}  |\widehat\psi^{g,M'^{-1}}|^2= \|P_g\psi\|^2$ in terms 
of orthogonal projectors $P_g$ onto the subspaces $\mathcal{V}_g$, and the  resolution of the identity
 $\sum_{g=0}^\infty P_g=I_{L^2(\mathbb{T}^2)}$. Since all $|\widehat\gamma^{g,g}|$ are greater than zero and uniformly bounded 
from above, we arrive to \eqref{Bessel1} with upper bound $C=\mathrm{max}_{g}\{|\widehat\gamma^{g,g}|^2\}$.$\blacksquare$

Proposition \ref{propBessel} provides an admissibility condition for modular ``coherent states''. Note that, in contrast to 
Proposition \ref{propGamma} and Theorem  \ref{continuousf}, now $\widehat\gamma$ does not need to have 
support on the four Fourier quadrants $Q_q, q=1,2,3,4$,  but only on the main diagonal $n_1=n_2$. 

The set $F_{\gamma}$ is not a frame in $L^2(\mathbb{T}^2)$, since $|\widehat\gamma^{g,g}|\rightarrow 0$ 
when $g\rightarrow \infty$, preventing $|\widehat\gamma^{g,g}|$ 
to be uniformly bounded from below by a positive constant. 
However if we restrict ourselves to suitable subspaces of $L^2(\mathbb{T}^2)$, like that of band-limited
functions 
\begin{equation}
W_{L_1,L_2}=\{\psi\in L^2(\mathbb T^2): \widehat\psi^{n_1,n_2}=0,\, \forall |n_1|>L_1, |n_2|>L_2\}\subset L^2(\mathbb T^2)\,,
\end{equation}
the set $F_{\gamma}$ becomes a frame, even for a suitable bandlimited function 
$\eta\in L^2(\mathbb{T}^1)$. More precisely, we have the following result:
\begin{corollary}
Under the conditions of Proposition \ref{modularBessel}, the set  $F_{\gamma}$ is a frame for any subspace $W_{L_1,L_2}$ of band 
limited functions  in $L^2(\mathbb{T}^2)$.
\end{corollary}
\textbf{Proof:} Let us consider the space of  band-limited functions of band-limits $L_1,L_2\in\mathbb N$ 
such that $\{0,1,\dots,g_{\rm max}\}\subset {\rm supp}(\widehat{\eta})$, where $g_{\rm max}={\rm max}(L_1,L_2)$. For functions
$\psi\in W_{L_1,L_2}$ $P_g\psi =0$ for $g>g_{\rm max}$, therefore the sum on $g$ in eq. \eqref{Bessel2} truncates and eq. \eqref{Bessel1} 
can be written as:
\begin{equation}
 c\|\psi\|^2< \int\frac{d\vartheta_1d\vartheta_2}{(2\pi)^2}\sum_{M\in{\cal X}} |\langle \gamma_M^{\vartheta_1,\vartheta_2}|\psi\rangle|^2 \leq  C ||\psi||^2\,,\,\,
\forall \psi\in W_{L_1,L_2}\,,
\end{equation}
where $c=\mathrm{min}_{g=0}^{g_{\rm max}}\{|\widehat\gamma^{g,g}|^2\}$ and  $C=\mathrm{max}_{g=0}^{g_{\rm max}}\{|\widehat\gamma^{g,g}|^2\}$.

Note that if $\gamma$ is chosen such that $\widehat{\eta}=\chi_{[0,g_\mathrm{max}]}$, then $F_\gamma$ is a tight frame, and a Parseval frame if
appropriately rescaled.

We believe that the frame property of $F_\gamma$ also holds for more general spaces of functions with rapidly decaying Fourier coefficients.


Next we combine the modular transformations and translations with diagonal dilations on the torus.

\subsection{Modular admissibility, modular wavelets and frame conditions} 
 We shall make use of the modular group 
to complete the parameter space $X'$ for the case of dependent dilations $a_2=\sigma(a_1)$ (for simplicity, we shall restrict
ourselves to the case $\sigma(a)=a$).  The action of the modular group on $\mathbb T^2$ induces a transformation of
functions $f\in L^2(\mathbb T^2)$ that completes the previous (dilation and translation) transformations as
\begin{equation}
 f_{a,M}^{\vartheta_1,\vartheta_2}(\theta_1,\theta_2):={f}_{a}^{\vartheta_1,\vartheta_2}
 (M^{-1}(\theta_1,\theta_2)^t)={f}_{a}^{\vartheta_1,\vartheta_2}
 (q\theta_1-n\theta_2,-p\theta_1+m\theta_2),
\end{equation}
where we have used the notation  ${f}_{a}^{\vartheta_1,\vartheta_2}:={f}_{a,a}^{\vartheta_1,\vartheta_2}$ when restricting to a single dilation
in equation \eqref{reptorus}, for convenience.

As we have seen in the previous section, adding the whole modular group $SL(2,\mathbb Z)$ to the parameter space $X'$ introduces redundancy that is not suitable
for admissibility conditions. Therefore, we shall restrict ourselves to the quotient space $\mathcal{X}=SL(2,\mathbb Z)/N$, where
$N$ refers to the isotropy subgroup \eqref{isotropydiag}.  The
choice $N$ (isotropy subgroup of $(g,g)$) is in fact connected with the case $\Gamma(\theta_1,\theta_2)=\eta(\theta_1+\theta_2)$, 
for which the only possible non-zero
Fourier coefficients are the diagonal $\widehat\Gamma^{l,l}$ (we shall make use of this property when proving the frame condition). 

The admissibility condition \eqref{admistrongT2} for ``modular wavelets'' on the torus\footnote{The term  ``modular wavelet''
was previously introduceced in \cite{Penn}, but in the rather different context of integral fractional linear transformations on the 
circle.}, can be restated as follows:

\begin{definition}
A non-zero function $\gamma\in L^2(\mathbb T^2)$ is called ``modular-admissible'' if there exist  $C\in\mathbb{R}$ such that 
the condition
\begin{equation}
0<\int_{X'} d\nu'(\vartheta_1,\vartheta_2,a)\sum_{M\in\cal X}|\langle {\gamma}_{a,M}^{\vartheta_1,\vartheta_2} |\psi\rangle|^2< C<\infty \label{admigeneralT2mod}
\end{equation}
is satisfied for every non-zero  $\psi \in L^2(\mathbb T^2)$.
\end{definition}

This admissibility condition can be equivalently expressed  as follows:

\begin{proposition}
  A non-zero function $\gamma\in L^2(\mathbb T^2)$ 
is ``modular-admissible'' iff there exist  $C\in\mathbb{R}$ such that
 \begin{equation}
 0<\widetilde\Lambda_{n_1,n_2}\equiv \int_0^\infty \frac{da}{a^3}\sum_{M\in\cal X}
|\widehat\gamma_{a,M}^{n_1,n_2}|^2< C < \infty\,,\forall (n_1,n_2)\in \mathbb{Z}^2 \label{admistrongT2mod}
 \end{equation}
where $\widehat\gamma_{a,M}^{n_1,n_2}=\langle \phi_{n_1,n_2}|\gamma_{a,M}\rangle$ are the Fourier
coefficients of $\gamma_{a,M}\equiv\gamma_{a,M}^{0,0}$.
\end{proposition}
\noindent {\bf Proof:} The proof follows similar steps as in Proposition \ref{admiprop1}. More precisely:
\begin{equation}
\int_{X'}  d\nu'(\vartheta_1,\vartheta_2,a)\sum_{M\in \mathcal X}
|\langle \gamma_{a_1,a_2}^{\vartheta_1,\vartheta_2}|\psi\rangle|^2
=\sum_{n_1,n_2=-\infty}^\infty\int_0^\infty\frac{da}{a^3}\sum_{M\in \mathcal X}
|\widehat{{\gamma}}^{n_1,n_2}_{a,M}|^2|\widehat{\psi}^{n_1,n_2}|^2,\label{uppermod}
\end{equation}
and this quantity is finite and non-zero if \eqref{admistrongT2mod} holds. $\blacksquare$ 

\begin{proposition}
The necessary admissibility condition \eqref{admiweak} still holds for modular admissible functions.
\end{proposition}
\noindent {\bf Proof:} Using the same reparametrization $(n_1,n_2)\sim (g,M^{-1})$ of the Fourier labels 
as in the proof of Proposition  \ref{propBessel}, we can write
 \begin{equation}
\widetilde\Lambda_{g,M'^{-1}}\equiv \widetilde\Lambda_{n_1,n_2}= \int_0^\infty \frac{da}{a^3}\sum_{M\in\cal X}
|\widehat\gamma_{a,M}^{g,M'^{-1}}|^2= \int_0^\infty \frac{da}{a^3}\sum_{M\in\cal X}
|\widehat\gamma_{a}^{g,M'^{-1}M}|^2\,\label{admimod}
\end{equation}
where we have denoted $\gamma_{a,I_2}=\gamma_a$ for simplicity. The approximation \eqref{smallscales} 
over small scales $a\ll 1$ can now be written as $\widehat\gamma_{a}^{g,M}\approx 2a\widehat\Gamma^{ag,M}$, and 
therefore it is again necessary that $\widehat\Gamma^{0,I_2}=0$, which is equivalent to \eqref{admiweak}.$\blacksquare$

Note that when writing $(ag,M)$, we are meaning $(an_1,an_2)=(\alpha_1,\alpha_2)$, which are not necessarily integers, but we preserve 
the ``modular information'' $(g,M)$ derived from $(n_1,n_2)$. Remember that $\widehat\Gamma^{n_1,n_2}$ can be extended to the 
reals $\widehat\Gamma^{\alpha_1,\alpha_2}$ in a continuous way, as commented in the proof of Theorem \ref{continuousf}. 

Without loss of generality, from now on 
we shall restrict ourselves  to ``diagonal'' functions $\Gamma(\theta_1,\theta_2)=\eta(\theta_1+\theta_2)$, 
for which $\widehat\Gamma^{n_1,n_2}=0$ if $n_1\not= n_2$, that is,  $\widehat\Gamma$ has only support on the main diagonal. 
Note that, introducing modular transformations relaxes the requirement that $\widehat\Gamma$ must have support on the four 
quadrants. Actually, it is just enough that  $\widehat\Gamma$ has support on the positive main diagonal, as it will be shown 
in the next Theorem.

\begin{theorem}\label{continuousfmod} For any localized modular-admissible function $\gamma$, whose 
associated function $\Gamma$ is diagonal, the family
\begin{equation}
 \left\{\gamma_{a,M}^{\vartheta_1,\vartheta_2},\ (\vartheta_1,\vartheta_2)\in(-\pi,\pi)^2, a\in(0,\infty),
 M \in\mathcal X\right\}
\end{equation}
is a frame, that is, there exist real constants $0<c\leq C$ such that
\begin{equation}
c||\psi||^2\leq \sum_{M\in \mathcal X}\int_{X'}  d\nu'(\vartheta_1,\vartheta_2,a)
|\langle \gamma_{a,M}^{\vartheta_1,\vartheta_2}|\psi\rangle|^2\leq C ||\psi||^2,\;\;\forall
\psi\in L^2(\mathbb T^2).\label{lowerupperbmod}
\end{equation}
\end{theorem}
\noindent{\bf Proof:}
It remains to prove the lower bound,  which is equivalent to prove that
$\widetilde\Lambda_{n_1,n_2}>c,\,\forall n_1,n_2\in\mathbb{Z}$. Following a similar strategy as in the proof of Theorem  \ref{continuousf}, 
we take $(\alpha_1,\alpha_2)=(\alpha^0,\alpha^0)$ such that $|\widehat\Gamma^{\alpha^0,\alpha^0}|>0$. By continuity, 
there exist $\rho$, with $0<\rho<|\alpha^0|$, such that 
$|\widehat\Gamma^{\alpha,\alpha}|>|\widehat\Gamma^{\alpha^0,\alpha^0}|/2$ in the interval
$(\alpha^0-\rho,\alpha^0+\rho)$. In \eqref{admimod} there will be values of $a$ and $M$ satisfying 
\begin{equation}
 a (n_1,n_2)M\simeq (\alpha^0,\alpha^0).
\end{equation}
Actually,  $M=M'=M_{n_1,n_2}$ in \eqref{representative} if $\alpha^0>0$, and $M=M'=-M_{n_1,n_2}$ if $\alpha^0<0$, 
and this means $a\simeq |\alpha^0|/g$. Therefore if we keep just this
term of the sum in \eqref{admimod} then we obtain:

 \begin{equation}
\widetilde\Lambda_{g,M'^{-1}}\equiv \widetilde\Lambda_{n_1,n_2}\geq \int_0^\infty \frac{da}{a^3}
|\widehat\gamma_{a}^{g,I_2}|^2\,.\label{admimod2}
\end{equation}
We shall consider the contribution to the integral \eqref{admimod2} that comes from the range
$a\in (\alpha^0-\rho,\alpha^0+\rho)/g$. Since $\gamma_a$ is integrable, its Fourier coefficients $\widehat\gamma_a^{n_1,n_2}$ 
tend to zero for $|n_1|,|n_2|\to\infty$, 
in particular $\widehat\gamma_a^{g,q}\to 0$ for $g\to\infty$. Therefore we need only to consider the 
less favorable case $g\gg 1$ implying $a\ll 1$.
Using the approximation \eqref{smallscales} for small $a_1=a_2=a$,  we can write
$\widehat{{\gamma}}^{g,g}_{a,I_2}\approx 2a \widehat{{\Gamma}}^{ag,ag}$ and
\begin{equation}
 \widetilde\Lambda_{n_1,n_2}\geq \int_{(\alpha^0-\rho)/g}^{(\alpha^0-\rho)/g}
\frac{da}{a}
4| \widehat{{\Gamma}}^{ag,ag}|^2> | \widehat{{\Gamma}}^{\alpha^0,\alpha^0}|^2\log\frac{\alpha_0+\rho}{\alpha_0-\rho},\label{frameboundmodmod}
\end{equation}
gives a strictly positive quantity independent of
$n_1,n_2$, which proves that
$\widetilde\Lambda_{n_1,n_2}$ is bounded from below. $\blacksquare$

\begin{figure}
\begin{center}
(a)\includegraphics[height=6cm]{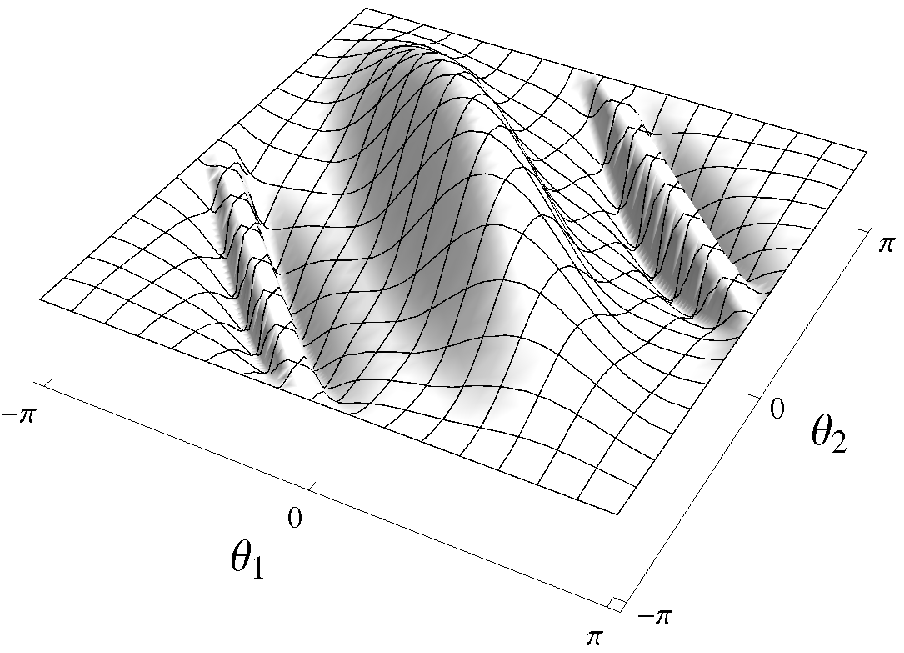}(b)\includegraphics[height=6cm]{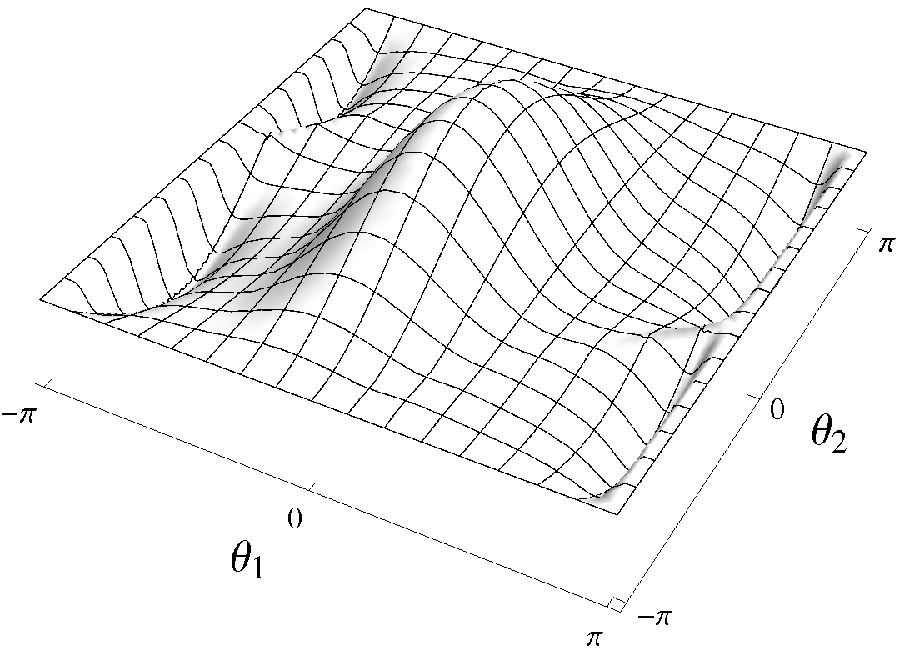}
(c)\includegraphics[height=6cm]{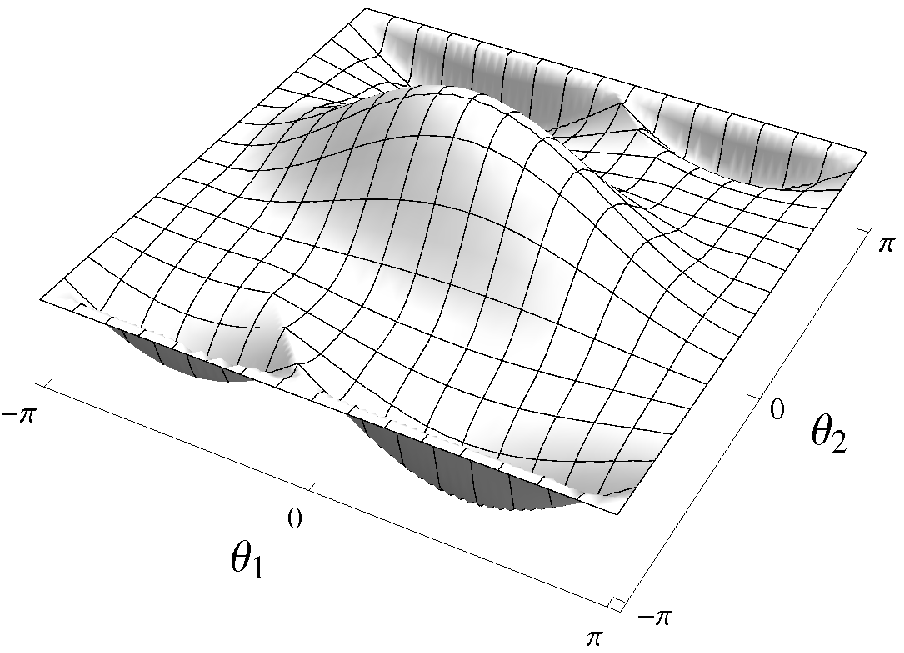}(d)\includegraphics[height=6cm]{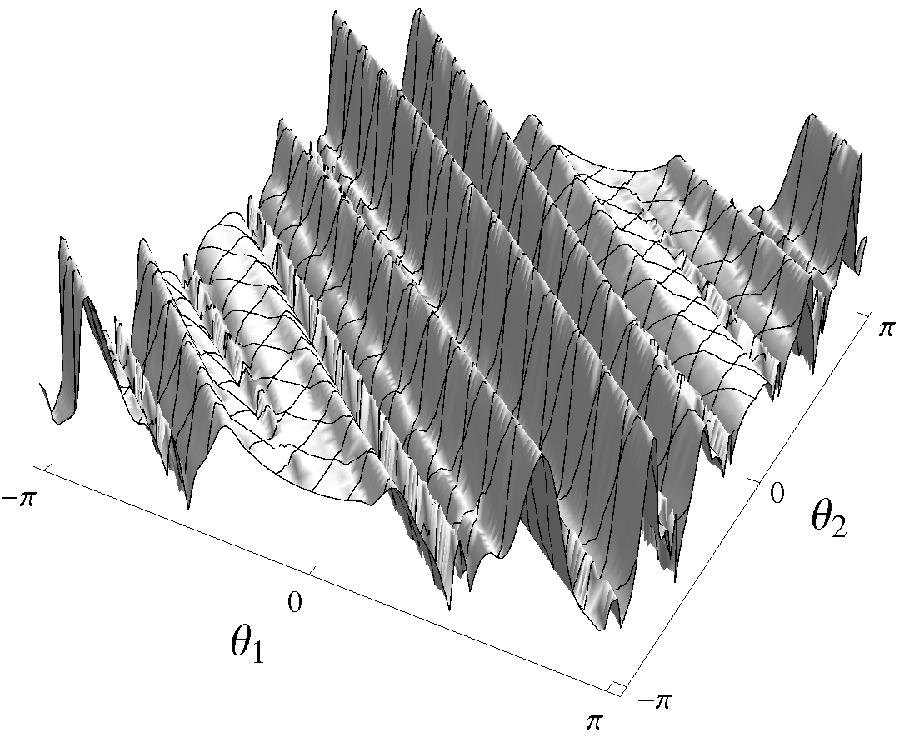}
\end{center}
\caption{ Modular transformation of the ``diagonal DoG'' \eqref{diagDoG} with $\alpha=10$ for: (a) $M=I_2$, (b) $M_{1,0}$, (c)  
$M_{0,1}$, (d)  $M_{4,5}$. }\label{DOGdiagmod}
\end{figure}

Let us provide a particular example of modular admissible function based on DoG functions \eqref{DoG}. Consider 
the diagonal function
\begin{equation}
 \Gamma(\theta_1,\theta_2)=\frac{\psi_\alpha\left(2\tan\frac{\theta_1+\theta_2}{2}\right)}{1+\cos(\theta_1+\theta_2)},
\end{equation}
so that the corresponding admissible function on the torus is the ``diagonal DoG''
\begin{equation}
 \gamma(\theta_1,\theta_2)= \sqrt{(1+\cos\theta_1)(1+\cos\theta_2)}\Gamma(\theta_1,\theta_2).\label{diagDoG}
\end{equation}
In Figure \ref{DOGdiagmod} we have plotted this function together with its modular transformation $\gamma_{M_{n_1,n_2}}$ for different 
values of $n_1, n_2$.

Analogous expressions for wavelet coefficients \eqref{wavecoef} and reconstruction formula \eqref{reconstruction} can be 
written for modular wavelets. 

\section{Conclusions}

In this article we have addressed the problem of constructing a CWT on the torus. 
Firstly we have derived the CWT on $\mathbb T^2$ entirely from the conformal group $SO(2,2)$. Proposition  \ref{propGamma} and 
Theorem \ref{continuousf}  yield the basic ingredients for writing a genuine
CWT on  $\mathbb T^2$ by  proving admissibility conditions and providing continuous frames and reconstruction 
formulas. The proposed CWT on  $\mathbb T^2$ 
has the expected Euclidean limit; that is, it behaves locally like the usual (flat) CWT on $\mathbb R^2$ but with two dilations 
(the natural tensor product representation of usual wavelets on $\mathbb R$). If one restricts oneself to a single (namely, diagonal) 
dilation, then the frame property is lost unless additional requirements on the support of $\widehat\gamma$ are imposed.
However, one can circumvent this problem by adding extra modular group $SL(2,\mathbb Z)$ transformations 
to the parameter space $X$ of the CWT, thus leading to the concept of \emph{modular wavelets}. Before defining modular-admissible 
functions and prove frame conditions in Theorem \ref{continuousfmod}, we have studied the modular group, its orbits in 
$\mathbb{Z}^2$, its unitary action  on $L^2(\mathbb T^2)$, invariant 
subspaces $\mathcal{V}_g\subset L^2(\mathbb T^2)$ and its orthonormal basis,  Bessel sequences and  modular frames for band limited functions.

In this article we have provided a CWT on the torus based on the theory of coherent states 
of quantum physics (formulated in terms of group representation theory). Another alternative construction 
based on area preserving projections for surfaces of revolution \cite{surfacesofrevolution} is the subject of another paper
in progress \cite{surfacetorus}.

Once we have studied the continuous approach, it remains to address the discretization, which roots
in the Littlewood-Paley analysis, and yields fast algorithms for computing the wavelet transform numerically. An
intermediate approach which paves the way between the continuous and the discrete cases is based on the representations
of some finite groups like in Ref. \cite{finitecircle} for wavelets on discrete fields (namely, the discrete circle 
$\mathbb Z_N=\mathbb Z/N\mathbb Z$). 

%

\section*{Acknowledgements}

We thank G. Garrig\'os for valuable discussions. This work was partially supported by the 
Spanish MICINN  (FIS2011-29813-C02-01), University of Granada (PP2012-PI04) and Fundaci\'on S\'eneca (08814/PI/08). 
D.R. was supported by the Sectorial Operational Programme Human Resources Development 2007-2013 of the 
Romanian Ministry of Labor, Family and Social Protection through the Financial Agreement  POSDRU/89/1.5/S/62557.


\begin{thebibliography}{99}

\bibitem{GMP} A. Grossmann, J. Morlet and T. Paul, Transforms associated
to square integrable group representations I. General results, J. Math.
Phys. \textbf{26} (1985) 2473-2479.

\bibitem{Gazeau} S.T. Ali, J-P. Antoine and J-P. Gazeau, Coherent States, Wavelets and Their
Generalizations, Springer (2000).


\bibitem{Fuhr} H. F\"uhr, Abstract Harmonic Analysis of Continuous Wavelet Transforms,
Springer Lecture Notes in Mathematics, vol. 1863,
Springer-Verlag, Heidelberg,
2005.


\bibitem{CWTmanifolds} J-P. Antoine, D. Ro\c sca, P. Vandergheynst, Wavelet transform on manifolds: Old and new
approaches, Appl. Comput. Harmon. Anal. \textbf{28} (2010) 189--202.

\bibitem{Fuhr2} H. F\"uhr, Painless Gabor expansions on homogeneous manifolds,
 Appl. Comput. Harmon. Anal. \textbf{26} (2009) 200--211.


\bibitem{Holschneider-sphere} M. Holschneider, Continuous Wavelet Transforms on the
sphere, J. Math. Phys. \textbf{37} (1996) 4156--4165.

\bibitem{waveS2} J-P. Antoine and P. Vandergheynst, Wavelets on the
2-sphere: a group-theoretical approach, Appl. Comput. Harmon.
Anal. \textbf{7}  (1999) 262--291.


\bibitem{nsphere}  J-P. Antoine and P. Vandergheynst, Wavelets on the
$n$-sphere and related manifolds, J. Math. Phys. \textbf{39}
(1998) 3987--4008.


\bibitem{Sphere-Implementation} J.-P. Antoine, L. Demanet, L. Jacques and P. Vandergheynst, 
Wavelets on the sphere: implementation and approximations
Appl. Comput. Harmon. Anal. \textbf{13} (2002) 177-200

\bibitem{AVdiscreteS2} I. Bogdanova, P. Vandergheynst, J-P. Antoine, L. Jacques, M. Morvidone:
Stereographic wavelet frames on the sphere,
Appl. Comput. Harmon. Anal. \textbf{19} (2005) 223-252.


\bibitem{cwthyperbol} I. Bogdanova, P. Vandergheynst and J-P. Gazeau, Continuous wavelet transform on the hyperboloid,
Appl. Comput. Harmon. Anal. \textbf{23} (2007) 285-306.


\bibitem{MacMahon} M. Calixto and E. P\'erez-Romero, Extended MacMahon-Schwinger's Master Theorem and 
Conformal Wavelets in Complex Minkowski Space, 
Appl. Comput. Harmon. Anal. \textbf{21} (2006) 204-229 

\bibitem{Prange} R. E. Prange and  S. M. Girvin, The Quantum Hall Effect, Springer London, Second Edition, (1990).

\bibitem{CMP} V. Aldaya, M.Calixto and J. Guerrero, Algebraic Quantization, Good Operators
and Fractional Quantum Numbers, Commun. Math. Phys. \textbf{178} (1996) 399-424 

\bibitem{modular} J. Guerrero, M. Calixto and V. Aldaya, Modular invariance on the torus and Abelian
Chern-Simons theory, J. Math. Phys. {\bf 40} (1999) 3773-3790


\bibitem{wavecircle} M. Calixto and  J. Guerrero,
Wavelet transform on the circle and the real line:
A unified group-theoretical treatment,
Appl. Comput. Harmon. Anal. {\bf 21} (2006) 204-229.

\bibitem{waveletmicro} R. Ashino, S.J. Desjardins, C. Heil, M. Nagase and R. Vaillancourt, 
Smooth tight frame wavelets and image microanalyis in the fourier domain, 
Comp. Math. Appl. \textbf{45} (2003) 1551-1579

\bibitem{toroidalmicro}
M. Ruzhansky and V. Turunen, Quantization of pseudo-differential operators on the torus, J. Fourier Anal. Appl. \textbf{16} (2010) 943-982

\bibitem{Barut} A.O. Barut and R. R\c{a}czka, {Theory of Group Representations and Applications},
Polish Scientific Publishers, Warszawa (1980).

\bibitem{Ole} O. Christensen, An introduction to frames and Riesz basis, Birkh\"auser, Boston (2003).



\bibitem{tensor} P. Wojtaszczyk, A mathematical introduction to wavelets, London 
Mathematical Society, Student Texts 37. Cambridge University Press 1997.


\bibitem{curveletrefs} E. J. Cand\`es and D. L. Donoho,
Continuous curvelet transform: I. Resolution of the wavefront set, Appl. Comput. Harmon. Anal. \textbf{19} (2005) 162-197\\
E. J. Cand\`es and D. L. Donoho, Continuous curvelet transform
II. Discretization and frames, Appl. Comput. Harmon. Anal. \textbf{19} (2005) 198-222.

\bibitem{shearletrefs} D. Labate, W.-Q Lim, G. Kutyniok and G. Weiss,
Sparse multidimensional representation using shearlets.
Wavelets XI (San Diego, CA, 2005), 254-262, SPIE Proc. 5914, SPIE, Bellingham, WA, (2005).

\bibitem{Gustavo} M. Frazier, G. Garrig\'os, K. Wang and G. Weiss, A characterization of functions that generate wavelet and
related expansion, J. Fourier Anal. Appl. \textbf{3} (1997)  883-906

\bibitem{weiss} D. Labate, G. Weiss and E. Wilson, Wavelets,
 Notices of the AMS {\bf 60} (2013) 66-76. 

\bibitem{orbit-stabilizer-theorem} John F. Humphreys, A course in group theory, Oxford University Press (1996).

\bibitem{Penn} R. C. Penner,  On Hilbert, Fourier, and Wavelet Transforms,
Comm. Pure Appl. Math. \textbf{55} (2002) 772-814


\bibitem{surfacesofrevolution} D. Ro\c sca, Wavelet analysis on some surfaces of revolution via area preserving
projection,
Appl. Comput. Harmon. Anal. {\bf 30} (2011) 262-272.


\bibitem{surfacetorus} M. Calixto, J. Guerrero and D. Ro\c sca, Wavelet transform on the torus: measure-preserving maps and
relation to the sphere, in progress.

\bibitem{finitecircle} 
K. Flornes, A. Grossmann, M. Holschneider, B. Torresani, Wavelets on discrete fields, Appl. Comput. Harmon. Anal. \textbf{1} (1994) 137.







\end{thebibliography}
\end{document}